\documentclass[aps,prd,superscriptaddress,longbibliography,nofootinbib,preprintnumbers,preprint]{revtex4-2}
\usepackage[utf8]{inputenc}
\usepackage[T1]{fontenc}
\usepackage{amsmath,amsfonts,amssymb,physics,subcaption}
\usepackage{mathtools}
\usepackage{mathrsfs}
\usepackage[usenames,dvipsnames]{xcolor}
\usepackage{setspace}
\usepackage{verbatim}
\usepackage{graphicx}
\usepackage{empheq}
\usepackage{pifont}
\usepackage{footmisc}
\usepackage{ragged2e}
\usepackage{caption}
\usepackage[colorlinks=true, linkcolor=RoyalBlue, citecolor=RoyalBlue, urlcolor=RoyalBlue]{hyperref}
\usepackage{footnotehyper}
\usepackage{float}

\usepackage{fancyhdr}
\def\f{\frac}
\def\mc{\mathcal}   
\def\mf{\mathfrak}

\def\v[#1]{\textbf{#1}}
\def\w[#1]{\widehat{#1}}
\def\vs[#1,#2]{\boldsymbol{{#1}_{#2}}}
\def\mes[#1]{d^{3}{#1}}
\def\del{\partial}
\def\vecs[#1,#2]{\boldsymbol{{#1}_{#2}}}

\newcommand{\be}{\begin{equation}}
\newcommand{\ee}{\end{equation}}

\def\a{\alpha}
\def\b{\beta}
\def\d{\delta}
\def\D{\Delta}
\def\e{\epsilon}

\def\g{\gamma}
\def\G{\Gamma}

\def\m{\mu}
\def\n{\nu}
\def\N{\nabla}

\def\O{\Omega}

\def\P{\Phi}
\def\vp{\varphi}
\def\s{\sigma}
\def\t{\tau}

\begin{document}

\title{Carroll hydrodynamics with spin}

\author{Ashish Shukla}
\email{ashish.shukla@saha.ac.in}
\affiliation{Theory Division, Saha Institute of Nuclear Physics, 1/AF Bidhan Nagar, Kolkata 700064, India.}
\affiliation{Homi Bhabha National Institute, Anushakti Nagar, Mumbai 400094, India.}

\author{Rajeev Singh}
\email{rajeev.singh@e-uvt.ro}
\affiliation{Department of Physics, West University of Timisoara, Bulevardul Vasile P\^{a}rvan 4, Timisoara 300223, Romania.}

\author{Pushkar Soni}
\email{pushkars21@iitk.ac.in}
\affiliation{Indian Institute of Technology Kanpur, Kanpur 208016, India.}

\begin{abstract}
We formulate Carroll hydrodynamics with the inclusion of a spin current. Our strategy relies on the fact that the $c\to 0$ limit of relativistic hydrodynamics yields the equations of Carroll hydrodynamics. Starting with the pre-ultralocal parametrization of the background geometry and the hydrodynamic degrees of freedom for a relativistic fluid endowed with a spin current, the $c\to 0$ limit produces Carroll hydrodynamics with spin. It is known that boost-invariant hydrodynamic models for ultrarelativistic fluids relevant for the physics of quark-gluon plasma, such as Bjorken and Gubser flow, are manifestations of Carroll hydrodynamics under appropriate geometric choices for the underlying Carrollian structure. In this work, we further this mapping between such boost-invariant models and Carroll hydrodynamics, now with the inclusion of a spin current.  
\end{abstract}


\maketitle
\thispagestyle{fancy}

\newpage
{\hypersetup{linkcolor=black}\tableofcontents}

\newpage
\section{Introduction}
\label{introduction}
Over the past decade, the study of heavy-ion collisions at the Relativistic Heavy Ion Collider (RHIC) has lead to an intriguing result. The experiments at RHIC have observed non-vanishing correlations between the angular momentum of the quark-gluon plasma (QGP) produced in off-centre heavy-ion collisions and the spin-polarization of the $(\Lambda, \bar{\Lambda})$-hyperons produced after the hadronization of the QGP \cite{STAR:2017ckg, STAR:2018gyt}. It has now been known for a while that the evolution of the QGP produced in heavy-ion collisions is well described by relativistic hydrodynamics \cite{Gale:2013da}. Such studies have in fact shown that QGP is the most ideal fluid in nature, with a shear viscosity to entropy density ratio of $\eta/s \approx 0.1 - 0.2$ just above the deconfinement temperature \cite{PhysRevC.104.054904}, which is very close to the conjectured Kovtun-Son-Starinets bound of $\eta/s \ge 1/4\pi$ that is met by all known fluids in nature \cite{Kovtun:2004de}. This immediately raises the question that can relativistic hydrodynamics be suitably modified with the addition of new degrees of freedom such that the observed non-vanishing correlations between the angular momentum of the QGP and the spin-polarization of $(\Lambda, \bar{\Lambda})$-hyperons can also be explained from a hydrodynamic perspective? 

To answer this question, there has been a sustained effort in the direction of developing what has now come to be known as ``relativistic spin hydrodynamics.'' Apart from the energy-momentum tensor and possible $U(1)$ currents, whose conservation equations correspond to the equations of relativistic hydrodynamics, one also introduces a ``spin current,'' which is a rank-three tensor, whose dynamics is expected to provide an explanation for the observed correlations between the angular momentum of the QGP and the spin-polarization of the $(\Lambda, \bar{\Lambda})$-hyperons \cite{Becattini:2018duy, Becattini:2020riu, Weickgenannt:2021cuo}. In fact, the earliest steps in the direction to relate the spin-polarization of hadrons produced in heavy-ion collisions and the vorticity of the QGP date back to \cite{Betz:2007kg, Becattini:2007sr}. Subsequent works have focused on understanding the emergence of spin-polarization from the distribution function of a many-body system comprising spin-$1/2$ particles \cite{Becattini:2013fla, Florkowski:2017ruc, Florkowski:2017dyn, Florkowski:2018ahw, Bhadury:2020puc, Bhadury:2020cop, Das:2022azr, Weickgenannt:2022qvh, Florkowski:2024bfw, Drogosz:2024gzv, Bhadury:2025boe}. An analysis based on the local second law of thermodynamics demanding positivity of entropy production has been carried out in \cite{Hattori:2019lfp, Biswas:2023qsw, Becattini:2023ouz, Daher:2025pfq, Becattini:2025oyi, Fang:2025aig}, while \cite{Gallegos:2020otk} attempts to provide a hydrodynamic explanation utilizing the anti de Sitter/conformal field theory (AdS/CFT) correspondence \cite{Maldacena:1997re}. In \cite{Abboud:2025shb}, relativistic spin hydrodynamics for an ideal fluid is formulated as a divergence-type theory, ensuring causality and stability of the dynamical equations, while semi-classical effects have been discussed in \cite{Chiarini:2024cuv}. A first-principles construction of relativistic spin hydrodynamics on backgrounds with intrinsic torsion has been attempted in \cite{Gallegos:2021bzp, Hongo:2021ona, Gallegos:2022jow}. Needless to say, relativistic spin hydrodynamics is presently a very active area of investigation.

One of the most prominent features of the heavy-ion collisions producing QGP is that the colliding nuclei are ultrarelativistic, moving almost at the speed of light. To gain analytic control over the complex physics of formation and subsequent spacetime evolution of QGP in such ultrarelativistic collisions, simple phenomenological models have been put forth in the past. These hydrodynamic models are built upon several simplifying assumptions, such as boost invariance of the flow along the beam axis, as well as rotation invariance in the transverse plane. The most celebrated of these models is Bjorken flow \cite{PhysRevD.27.140}, which, apart from the above symmetries, also assumes translation invariance in the transverse plane. This restricts the energy density profile of the QGP to depend only upon the proper time elapsed since the collision. Subsequent work relaxed the highly restrictive requirement of translation invariance with invariance under a combination of translations and special conformal translations in the transverse plane, known as Gubser flow \cite{PhysRevD.82.085027, Gubser:2010ui}, allowing for energy density profiles which are functions of both proper time and the radial distance away from the beam axis, albeit valid only for conformal fluids. These models have played a crucial role in our understanding of the physics of QGP, especially at early times after local thermal equilibrium has set in. In particular, recent work has lead to the extension of these models to also include a spin current \cite{Florkowski:2019qdp, Singh:2020rht, Wang:2021wqq}, in order to serve as simple analytic templates for spin transport in relativistic spin hydrodynamics.

An interesting development lately has been the realization that Bjorken and Gubser flow models are examples of Carroll hydrodynamics \cite{Bagchi:2023ysc, Bagchi:2023rwd}. Carrollian physics arises in the $c\to 0$ limit of Lorentzian dynamics \cite{levy1965nouvelle, SenGupta:1966qer}.\footnote{See \cite{Bagchi:2025vri} for a review of Carrollian physics. Ref.~\cite{Nguyen:2025zhg} also provides a review focusing on holographic aspects.} In this limit, the Poincar\'{e} algebra contracts to give rise to the Carroll algebra, which has several distinguishing features. These include the Hamiltonian becoming a central element of the algebra, while Carroll boosts commute with one-another unlike their Lorentzian counterparts. From an intrinsic perspective, the Carroll limit makes the dynamics ultralocal, as local lightcones collapse when $c\to 0$. Thus, motion along spatial directions becomes forbidden. On the other hand, from the perspective of an extrinsic observer, the findings of \cite{Bagchi:2023ysc, Bagchi:2023rwd} solidified the idea that the Carroll limit can also be thought of as an ultrarelativistic limit. Since non-relativistic Galilean physics of everyday experience can be arrived at from Lorentzian physics by taking the speed of light to infinity, $c\to\infty$, it can therefore intuitively be expected that the opposite limit, $c\to 0$, should lead to ultrarelativistic dynamics. Equations of Carroll hydrodynamics, which describe a Carrollian fluid evolving on a degenerate Carroll manifold,\footnote{See the review \cite{Ciambelli:2025unn} for a detailed discussion of Carrollian geometry.} can be arrived at by imposing the limit $c\to 0$ on the equations of relativistic hydrodynamics \cite{Ciambelli:2018xat, Petkou:2022bmz}.\footnote{See \cite{Armas:2023dcz} for an alternate construction of Carroll hydrodynamics, starting from an equilibrium generating functional and systematically accounting for the dynamics of the Goldstone bosons associated with the spontaneous breaking of Carroll boost symmetry. This leads to two distinct classes of Carroll fluids, of which one class corresponds to the Carroll fluids obtained from the $c\to 0$ limiting procedure from relativistic hydrodynamics, pertinent to the present discussion.} It was found in \cite{Bagchi:2023ysc, Bagchi:2023rwd} that for specific choices of  the geometric data for the background Carroll structure, the equations of Carroll hydrodynamics became identical to the equations governing Bjorken and Gubser flow. This mapping is in fact one of the prominent instances where the abstract structures of Carrollian physics make direct contact with more familiar setups,\footnote{Other important applications of Carrollian physics to real world systems include connections with the physics of fractons \cite{Bidussi:2021nmp}, flat bands \cite{Bagchi:2022eui, Ara:2024fbr}, waves in shallow water \cite{Bagchi:2024ikw}, phase separation in Luttinger liquids \cite{Biswas:2025dte}, and cosmology \cite{deBoer:2021jej}.} and was subsequently exploited in \cite{Kolekar:2024cfg} to compute departures from the strict phenomenological assumptions of the Bjorken and Gubser flow models, by retaining subleading terms in a systematic $c\to 0$ expansion of relativistic hydrodynamics. 

In the present work, motivated by the above mentioned relation between Carroll hydrodynamics and boost-invariant models of heavy-ion collisions, as well as the fact that the generalizations of these models that include a spin current can be relevant for understanding the observed spin polarization effects, we embark upon constructing the formalism for Carroll hydrodynamics with a spin current. The strategy for doing so is straightforward - we start from the equations of relativistic spin hydrodynamics and impose the $c\to 0$ limit on them. To do so consistently in a covariant manner, we first express the background geometry and fluid variables in a pre-ultralocal (PUL) parametrization \cite{Hansen:2021fxi, Armas:2023dcz}, which is adapted for the Carrollian structure to emerge naturally in the limit $c\to 0$. Further, in this paper, our focus is on ideal Carroll fluids, implying that the constitutive relations for the energy-momentum tensor and the spin current we employ are devoid of derivative corrections. The final result of this process is a set of hydrodynamic equations for an ideal Carroll fluid endowed with a spin current on a general Carrollian manifold. To further strengthen the mapping between Carroll hydrodynamics and boost-invariant models of heavy-ion collisions, we also work out the appropriate choices for the Carroll geometric and fluid data that maps the equations of Carroll hydrodynamics with a spin current to the generalizations of Bjorken and Gubser flow models that have a spin current. Apart from connections to the physics of QGP, we hope that the initiation of the study of Carroll hydrodynamics with spin currents will open up several new interesting avenues for exploration, including potential connections with condensed matter systems as well, where spin hydrodynamics may lead to important observational effects \cite{Jaiswal:2024urq}.

The paper is organized as follows. In section \ref{main_meat}, we establish the formalism for Carroll hydrodynamics with a spin current, for an ideal Carroll fluid. In section \ref{boost_inv_models}, we review the Bjorken and Gubser flow models, with the inclusion of a spin current, and exhibit the geometric choices that map ideal Carroll fluids with a spin current to these boost-invariant models for the dynamics of the QGP. Section \ref{discussion} concludes the paper with a discussion and an outlook towards problems that might be of interest for future work. Appendices \ref{full_terms} - \ref{pr_gauge} provide additional material relevant for the discussion in the main text.

\medskip

\noindent \underline{\textbf{Notation}}: We work in four dimensional spacetime, where the Lorentzian geometry carries the signature $(-,+,+,+)$, while its degenerate Carrollian counterpart has the signature $(0,+,+,+)$. Greek letters $\mu,\nu,\ldots$ in the superscript/subscript denote spacetime indices. Latin letters $A,B,\ldots$ denote tangent space indices, while $a,b,\ldots$ denote tangent space spatial indices. The flat space Minkowski metric and its inverse are denoted by $\eta_{AB}$ and $\eta^{AB}$, respectively. Symmetrization and antisymmetrization of indices follows the convention $A_{(\m} B_{\n)} \equiv \f{1}{2} (A_\m B_\n + A_\n B_\m)$ and $A_{[\m} B_{\n]} \equiv \f{1}{2} (A_\m B_\n - A_\n B_\m)$.

\section{Carroll hydrodynamics and spin currents}
\label{main_meat}
The following subsections detail the construction of hydrodynamics for a Carroll fluid endowed with a spin current. In subsection \ref{rel_hydro} we first provide a brief overview of the equations of relativistic hydrodynamics with a spin current. This is followed by a discussion on the pre-ultralocal (PUL) parametrization for the background geometry and the fluid degrees of freedom, which provides a covariant way to impose the $c\to 0$ limit on relativistic hydrodynamics, in subsection \ref{pul_setup}. In subsection \ref{carroll_spin_hydro}, we report the equations governing an ideal Carroll fluid with spin. Appendix \ref{full_terms} contains further details on various terms in the PUL parametrization of relativistic hydrodynamics which contribute only subleading pieces in a $c\to 0$ expansion, while appendix \ref{symmetries} discusses the local Carroll boost invariance of the Carroll hydrodynamic equations with spin. Appendix \ref{pr_gauge} presents an alternate non-covariant approach to arrive at Carroll hydrodynamics with spin, based on the Papapetrou-Randers (PR) parametrization of the background geometry and the fluid degrees of freedom. 

\subsection{Relativistic hydrodynamics with a spin current}
\label{rel_hydro}
We begin the discussion with an overview of relativistic hydrodynamics with a spin current. Our focus will be on ideal relativistic fluids. In other words, we are looking at the hydrodynamic regime of field theories with Poincaré invariance in thermal equilibrium. In flat Minkowski spacetime, application of the Noether procedure for spacetime translation and Lorentz invariance yields conservation equations for the energy-momentum and the total angular momentum currents of the system,
\be
\del_\m T^{\m\n} = 0\, ,\qquad  \del_\m J^{\m\n\lambda} = 0\, ,
\label{cons_eq_1}
\ee
where $T^{\m\n}$ is the energy-momentum tensor, while $J^{\m\n\lambda}$ is the total (i.e.~orbital plus spin) angular momentum tensor, given by
\be
J^{\m\n\lambda} = x^\n T^{\m\lambda} - x^\lambda T^{\m\n} - S^{\m\n\lambda}.
\label{total_ang_mom}
\ee
Here $S^{\m\n\lambda}$ denotes the spin angular momentum tensor, more commonly referred to as the ``spin current.'' Note that the angular momentum and spin currents are antisymmetric in the last two indices i.e.~$J^{\m\n\lambda} = - J^{\m\lambda\n}, S^{\m\n\lambda} = - S^{\m\lambda\n}$. Combining the decomposition eq.~\eqref{total_ang_mom} with the conservation laws eq.~\eqref{cons_eq_1} immediately yields
\be
\del_\m S^{\m\n\lambda} = T^{\n\lambda} - T^{\lambda\n}.
\label{spin_non_cons}
\ee
Thus, the non-conservation of spin current is related to the antisymmetric part of the energy-momentum tensor, which may be non-zero, as the Noether procedure does not guarantee a symmetric energy-momentum tensor. However, by adding a Belinfante-Rosenfeld (BR) improvement term to the energy-momentum tensor, it can be made symmetric. The improved BR energy-momentum tensor is given by
\be
T^{\m\n}_{\rm BR} \equiv T^{\m\n} - \frac{1}{2} \, \del_\lambda\!\left(S^{\lambda\m\n} - S^{\m\lambda\n} - S^{\n\lambda\m}\right).
\ee
It is straightforward to check that $T^{\m\n}_{\rm BR} = T^{\n\m}_{\rm BR}$. Further, $T^{\m\n}_{\rm BR}$ is conserved, $\del_\m T^{\m\n}_{\rm BR} = 0$. In terms of the improved energy-momentum tensor, the conserved total angular momentum tensor just becomes $J^{\m\n\lambda} = x^\n T^{\m\lambda}_{\rm BR} - x^\lambda T^{\m\n}_{\rm BR}$, up to a total derivative term, thereby removing the spin current entirely from the description.\footnote{When constructing the conserved charges i.e.~the Lorentz generators $M^{\m\n}$ from the current $J^{\m\n\lambda}$, via $M^{\m\n} = \int_\Sigma J^{0\m\n}$, the total derivative terms will not give any contribution on a manifold $\Sigma$ without boundary.} Thus, the BR improvement procedure highlights that the total angular momentum obtained using the Noetherian approach cannot be decomposed unambiguously into an orbital and a spin part. More generally, one has what is known as the ``pseudogauge freedom,'' which allows one to redefine the Noetherian energy-momentum tensor and the spin current via
\be
\begin{split}
&T^{\m\n} \rightarrow \tilde{T}^{\m\n} = T^{\m\n} - \frac{1}{2} \, \del_\lambda\!\left(Z^{\lambda\m\n} - Z^{\m\lambda\n} - Z^{\n\lambda\m}\right) , \\
&S^{\m\n\lambda} \rightarrow \tilde{S}^{\m\n\lambda} = S^{\m\n\lambda} - Z^{\m\n\lambda}\, , 
\end{split}
\label{pseudogauge}
\ee
where $Z^{\m\n\lambda}$ is an arbitrary rank-three tensor, with $Z^{\m\n\lambda} = -Z^{\m\lambda\n}$. Under the pseudogauge transformation eq.~\eqref{pseudogauge}, the total angular momentum $J^{\m\n\lambda}$ remains unchanged up to a total derivative term. Thus, physical quantities remain unaltered under a pseudogauge transformation, highlighting the ambiguity in decomposing the total angular momentum into an orbital and a spin part. In particular, the BR improvement procedure is a specific choice for the pseudogauge, namely the one corresponding to $Z^{\m\n\lambda} = S^{\m\n\lambda}$.

An unambiguous method to obtain the energy-momentum tensor and the spin current is to couple the field theory to a background geometry that carries intrinsic torsion \cite{Gallegos:2021bzp, Hongo:2021ona, Gallegos:2022jow}. The energy-momentum tensor is then sourced by the vielbein field, whereas the spin current is sourced by the spin connection of the background. Since the vielbein field and the spin connection are independent geometric data when the background is endowed with intrinsic torsion, the resulting energy-momentum and spin currents they source can not be reshuffled and absorbed into one another, in contrast to the flat space Noetherian approach. The background is further assumed to have a timelike Killing vector field, which characterizes thermal equilibrium. The requirements of diffeomorphism and local Lorentz invariance for the equilibrium generating functional \cite{Banerjee:2012iz, Jensen:2012jh, Kovtun:2016lfw, Kovtun:2018dvd, Shukla:2019shf, Kovtun:2019wjz} then yield the following conservation equations for the energy-momentum and spin currents,\footnote{Note that eqs.~\eqref{hydro_eqs} and \eqref{const_rels_1} have been written using a different derivative counting scheme compared to \cite{Gallegos:2021bzp, Hongo:2021ona, Gallegos:2022jow}. The scheme here is more in line with the usual counting, where an object with $n$ derivatives is counted as $n$-th order, while \cite{Gallegos:2021bzp, Hongo:2021ona, Gallegos:2022jow} work with an energy-momentum tensor with an antisymmetric part, where terms carrying $n$ derivatives are counted as $(n+1)$-th order in the derivative expansion.}
\be
\N_\mu T^{\mu\nu} = 0\, , \quad \N_\m S^{\m\n\lambda} = 0\, ,
\label{hydro_eqs}
\ee
where the covariant derivatives are with respect to the Levi-Civita connection, and the ideal fluid constitutive relations are\footnote{Approaches based specifically on kinetic theory \cite{DeGroot:1980dk, Florkowski:2018ahw} advocate for more terms in the constitutive relation for the spin current in eq.~\eqref{const_rels_1}. Our discussion will be restricted to the terms that are common and arise universally in various approaches to relativistic spin hydrodynamics.}
\be
\begin{split}
T^{\m\n} &= \e\, \f{u^\m u^\n}{c^2} + P \D^{\m\n}\, ,\\
S^{\m\n\lambda} &= u^\m \O^{\n\lambda}\, .
\label{const_rels_1}
\end{split}
\ee
Note that the background torsion has been turned off in eqs.~\eqref{hydro_eqs} and \eqref{const_rels_1} after performing the variation and obtaining the currents. In the constitutive relations above, $\e, P$ and $\O^{\m\n}$ (satisfying $\O^{\m\n} = - \O^{\n\m}$) respectively denote the energy density, pressure and spin density (sometimes also refereed to as the spin polarization tensor) for the fluid, which are functions of the hydrodynamic degrees of freedom viz.~the temperature $T$, the fluid four-velocity $u^\m$ (normalized such that $u^\m u_\m = -c^2$), and the spin chemical potential $\m^{\a\b}$ (satisfying $\m^{\a\b} =-\m^{\b\a}$). Also, $\D_{\m\n} \equiv g_{\m\n} +\frac{u_\m u_\n}{c^2}$ is the projector orthogonal to fluid four-velocity $u^\m$.
It turns out to be convenient for later use to carry out an electric-magnetic like decomposition of the spin density via
\be
\O^{\m\n} \equiv u^\m \P^\n - u^\n \P^\m + \e^{\m\n\a\b} u_\a \Pi_\b\, ,
\label{omega_decomp}
\ee
where $\e^{\m\n\a\b}$ is the Levi-Civita tensor,\footnote{$\e^{\m\n\a\b} = \varepsilon^{\m\n\a\b}/\sqrt{-g}$, with $\varepsilon^{\m\n\a\b}$ being the totally antisymmetric Levi-Civita symbol, with $\varepsilon^{0123} = +1$.} and the components $\P^\m, \Pi^\m$ are orthogonal to the fluid velocity i.e.~$u\cdot\P = u\cdot\Pi = 0$. Thus $\P^\m, \Pi^\m$ each carries three degrees of freedom, totaling the six degrees of freedom in the spin density $\O^{\m\n}$. The decomposition eq.~\eqref{omega_decomp} can also be inverted to give
\be
\P^\m = \f{1}{c^2} u_\n \O^{\m\n}\, , \quad \Pi^\m = - \f{1}{2c^2} \epsilon^{\m\n\a\b} u_\n \O_{\a\b}.
\ee
In the subsequent discussion, we will directly impose the Carroll limit on fields $\P^\m$ and $\Pi^\m$.

\subsection{Pre-ultralocal (PUL) parametrization and the $c\to 0$ limit}
\label{pul_setup}
To obtain the equations of Carroll hydrodynamics with a spin current, let us now discuss the pre-ultralocal (PUL) parametrization for the background Lorentzian geometry and fluid degrees of freedom, on which the $c\to 0$ limit is to be imposed. The discussion in this subsection closely follows \cite{Hansen:2021fxi, Armas:2023dcz}. As mentioned earlier, the PUL parametrization offers the advantage of being fully covariant, where imposing the $c\to 0$ limit lands us naturally upon an ultralocal Carrollian structure without referring to any specific choice of coordinates.\footnote{One may contrast this with the Papapetrou-Randers (PR) parametrization of a Lorentzian geometry, which relies on using a specific set of coordinates that bring the Lorentzian metric into a form suitable for a $3+1$-like split, before imposing the $c\to 0$ limit. The PR parametrization was used in previous works on Carroll hydrodynamics \cite{Ciambelli:2018xat, Petkou:2022bmz, Bagchi:2023ysc, Bagchi:2023rwd, Kolekar:2024cfg}, and its usage for incorporating spin currents in Carroll hydrodynamics is discussed in appendix \ref{pr_gauge} of the present work.} Additionally, the mapping of local Lorentz symmetries in the tangent space to Carrollian symmetries when $c\to 0$ is also transparent in the PUL parametrization. One begins by expressing the Lorentzian vielbein fields $E_\mu^{~\,A}$ and their inverses $E^\mu_{~A}$ as \cite{Hansen:2021fxi}
\be
E_\mu^{~\,A} = (c L_\mu, E_\mu^{~a})\, , \quad E^\mu_{~A} = \left(-\frac{1}{c} K^\mu, E^\mu_{~a}\right).
\label{pul_vielbeins}
\ee
As is evident from above, the tangent space timelike parts of the vielbein and its inverse viz.~$L_\mu, K^\mu$ have been written out explicitly, after extracting out factors of $c$.\footnote{If one chooses to work with coordinates $(t, x^i)$, this implies that the indices $\m, \n, \ldots$ when specialized to the time coordinate now become $t$ instead of the numeral $0$. The extracted factor of $``c"$ comes basically by trading $x^0$  for $c t$.} The Lorentzian metric $g_{\mu\nu} = E_\mu^{~\,A} E_\nu^{~\,B} \eta_{AB}$ and its inverse $g^{\mu\nu} = E^\mu_{~A} E^\nu_{~B} \eta^{AB}$ are then given by
\be
g_{\mu\nu} = - c^2 L_\m L_\n + H_{\m\n}\, , \quad g^{\m\n} = - \frac{1}{c^2} K^\m K^\n + H^{\m\n},
\label{metric_pul}
\ee
where $H_{\m\n} \equiv E_\m^{~a} E_\n^{~b} \d_{ab}$ and $H^{\m\n} \equiv E^\m_{~a} E^\n_{~b} \d^{ab}$. The indices on $H_{\m\n}$ can be raised using the inverse metric, while the ones on $H^{\m\n}$ can be lowered using the metric. One further has the relations $K^\m = c^2 g^{\m\n} L_\n$ and $L_\m = \frac{1}{c^2} g_{\m\n}K^\n$. The objects $K^\m, L_\m, E_\m^{~a}, E^\m_{~a}, H_{\m\n}$ and $H^{\m\n}$ are termed as the ``PUL variables.'' They admit the following orthogonality and completeness relations,
\be
\label{completeness_1}
K^\m L_\m = -1\, , \quad K^\m H_{\m\n} = L_\m H^{\m\n} = 0\, ,\quad  H^{\m\n} H_{\n\lambda} - K^\m L_\lambda = \d^\m_\lambda\, .
\ee

The next step is to take the following ansatz for the Carroll limit of the PUL variables, justified under the assumption of analyticity in the $c\to 0$ limit,
\be
\begin{split}
&K^\m = k^\m + \mc{O}(c^2)\, , \quad L_\mu = \ell_\mu + \mc{O}(c^2)\, , \quad E_\m^{~a} = e_\m^{~a} + \mc{O}(c^2)\, , \quad E^\m_{~a} = {e}^\m_{~a} + \mc{O}(c^2)\, , \\
&H_{\m\n} = h_{\m\n} + \mc{O}(c^2)\, , \quad H^{\m\n} = h^{\m\n} + \mc{O}(c^2)\, .
\end{split}
\label{PUL_Carr_Exp}
\ee
Here $h_{\m\n} \equiv e_\m^{~a} e_\n^{~b} \d_{ab}$ and $h^{\m\n} \equiv {e}^\m_{~a} {e}^\n_{~b} \d^{ab}$. Further, the orthogonality and completeness relations eq.~\eqref{completeness_1} now imply
\be
\label{completeness_2}
k^\m \ell_\m = -1\, , \quad k^\m h_{\m\n} = \ell_\m h^{\m\n} = 0\, ,\quad  h^{\m\n} h_{\n\lambda} - k^\m \ell_\lambda = \d^\m_\lambda\, .
\ee
The advantage of working with the PUL variables is now evident, as we have arrived at the data defining a Carrollian structure without recourse to any specific choice of coordinates on the background geometry. The degenerate spatial metric $h_{\m\n}$ with the signature $(0,+,+,+)$ and its kernel $k^\mu$ give rise to a (weak) Carrollian structure \cite{Henneaux:1979vn, Duval:2014uoa}. The one-form field $\ell_\mu$ is the clock-form, while $h^{\m\n}$ is the co-metric of the Carrollian structure. 

It is also straightforward to see how local Lorentz transformations map to Carroll symmetries from the PUL perspective. To wit, under a local Lorentz transformation, denoted by $\Lambda^A_{~\,B}$, such that $\Lambda^{AB} = - \Lambda^{BA}$, the Lorentzian vielbein and its inverse transform as
\be
\d_\Lambda E_\m^{~\,A} = \Lambda^A_{~\,B}  E_\m^{~\,B}\, , \quad \d_\Lambda E^\m_{~A} = - \Lambda^B_{~\,A}  E^\m_{~B}\, , 
\ee
which keeps the Minkowski metric $\eta_{AB}$ and its inverse $\eta^{AB}$ invariant. In terms of the PUL variables, the local Lorentz transformations act via 
\be
\begin{split}
& \d_\Lambda K^\m = c\Lambda^a_{~0} E^\m_{~\,a}\, , \quad \d_\Lambda L_\m = \frac{1}{c} \Lambda^0_{~a} E_\m^{~a}\, , \\
& \d_\Lambda E_\m^{~\,a} = c\Lambda^a_{~0} L_\m + \Lambda^a_{~b} E_\m^{~b}\, , \quad \d_\Lambda E^\m_{~\,a} = \frac{1}{c} \Lambda^0_{~a} K^\m - \Lambda^b_{~a} E^\m_{~\,b}\,. 
\end{split}
\label{LLT_exp}
\ee
Now, under the $c\to 0$ limit, the spatial rotations and Lorentz boosts become
\be
\Lambda^a_{~\,b} = \lambda^a_{~\,b} + \mc{O}(c^2)\, , \quad \Lambda^0_{~\, a} = c \, \lambda_a + \mc{O}(c^3)\, , \quad \Lambda^a_{~\,0} = c \, \lambda^a + \mc{O}(c^3)\, ,
\ee
with $\lambda^a_{~\,b}$ and $\lambda_a$ respectively corresponding to spatial rotations and local Carroll boosts, under which, using eqs.~\eqref{PUL_Carr_Exp} and \eqref{LLT_exp}, the entities defining the Carrollian structure transform via
\be
\d_\lambda k^\mu = 0\, , \quad \d_\lambda \ell_\mu = \lambda_a e_\mu^{~\,a}\, , \quad \delta_\lambda e_\mu^{~\,a} = \lambda^a_{~\,b} e_\mu^{~\,b}\, , \quad \d_\lambda {e}^\mu_{~a} = \lambda_a k^\mu - \lambda^b_{~a} {e}^\mu_{~b}\, .
\label{lcb_trans}
\ee
In particular, under local Carroll boosts, while $(h_{\mu\nu}, k^\mu)$ are invariant, the clock-form and the co-metric shift via 
\be
\d_\lambda \ell_\mu = \lambda_a e_\mu^{~a} \equiv \lambda_\m \, , \quad \d_\lambda h^{\m\n} = \lambda^a ({e}^\mu_{~\,a} k^\nu + {e}^\nu_{~\,a} k^\mu) \equiv \lambda^\m k^\n + \lambda^\n k^\m\, ,
\label{lcb_trans_2}
\ee
where we have introduced the notation $\lambda_\mu \equiv \lambda_a e_\mu^{~a}$ and $\lambda^\mu \equiv \lambda^a {e}^\m_{~a}$, satisfying $\lambda_\m = h_{\m\n} \lambda^\n$ and $\lambda^\m = h^{\m\n} \lambda_\n$. Thus local Carroll boosts highlight a gauge redundancy in the co-metric and the clock-form $(h^{\m\n}, \ell_\m)$ on a Carrollian structure, and physical quantities should be constructed from boost-invariant combinations of them. 

The weak Carrollian structure that emerges from the $c\to 0$ limit of the PUL variables can be imbued with a Carroll-compatible connection, leading it to become a ``strong'' Carroll structure. A Carroll-compatible connection must preserve the spatial metric $h_{\m\n}$ and the kernel $k^\m$ that generates the null hypersurface, as these comprise the local Carroll boost invariant objects determining the Carrollian structure. Thus, one demands
\be
\widehat{\N}_\m k^\n = 0\, , \quad \widehat{\N}_\m h_{\n\lambda} = 0 .
\label{connection_req}
\ee
One need not demand the compatibility of the connection with the clock form $\ell_\mu$ or the co-metric $h^{\m\n}$, as they shift under local Carroll boosts. Following the discussion in \cite{Bekaert:2015xua, Hartong:2015xda, Hansen:2021fxi}, the conditions in eq.~\eqref{connection_req} are not sufficient to uniquely fix the connection. A useful choice is given by \cite{Hansen:2021fxi}
\be
\widehat{\G}^\m_{\n\lambda} = - k^\m \!\left(\del_{(\n} \ell_{\lambda)} + \ell_{(\n} \pounds_k \ell_{\lambda)}\right) +\f{1}{2}\,h^{\m\rho} \left(\del_\n h_{\rho\lambda} + \del_\lambda h_{\rho\n} - \del_\rho h_{\n\lambda}\right) - h^{\m\rho} \ell_\lambda \widehat{\mc{K}}_{\n\rho}\, .  
\label{carr_comp_conn}
\ee
Here $\pounds_k$ denotes the Lie derivative with respect to $k^\m$, and $\widehat{\mc{K}}_{\m\n} \equiv -\f{1}{2} \pounds_k h_{\m\n}$ is the extrinsic curvature determining how the spatial sections evolve along the null direction for a chosen foliation of the Carrollian structure. Note that $\widehat{\mc{K}}_{\m\n}$ is purely spatial in nature, as $k^\m \widehat{\mc{K}}_{\m\n} = 0$. Interestingly, under a local Carroll boost transformation eq.~\eqref{lcb_trans}, $\widehat{\G}^\m_{\n\rho}$ is not invariant, but rather changes by
\be
\d_\lambda\widehat{\G}^\m_{\n\rho} = -k^\m \!\left[\pounds_k\!\left(\lambda_{(\n} \ell_{\rho)}\right) +\ell_\rho \lambda^\s \widehat{\mc{K}}_{\s\n} + \f{1}{2} \pounds_\lambda h_{\n\rho}\right] + 2h^{\m\s} \lambda_{[\s} \widehat{\mc{K}}_{\rho]\n}\, .
\label{conn_var}
\ee
However, the conditions in eq.~\eqref{connection_req} remain invariant under a local Carroll boost because the changes to their left hand side vanish, $k^\rho \d_\lambda\widehat{\G}^\m_{\n\rho} = 0$ and $h_{\m(\s} \d_\lambda\widehat{\G}^\m_{\n\rho)} = 0$.

It is important to note that $\widehat{\G}^\m_{\n\lambda}$ carries torsion, and therefore cannot be arrived at by the $c\to 0$ limit of the Levi-Civita connection $\G^{\m}_{\n\lambda}$ in the parent Lorentzian geometry. One way to arrive at the connection $\widehat{\G}^\m_{\n\lambda}$ is to start with a non-Levi-Civita connection on the parent Lorentzian geometry, which satisfies similar requirements as eq.~\eqref{connection_req} in terms of the PUL variables before the $c\to 0$ limit is imposed, 
\be
\widetilde{\N}_\m K^\n = 0\, , \quad \widetilde{\N}_\m H_{\n\lambda} = 0\, .
\ee
A possible choice for this PUL connection that reduces to the Carroll-compatible connection $\widehat{\G}^\m_{\n\lambda}$ when the $c\to 0$ limit is imposed is
\cite{Hansen:2021fxi}
\be
\widetilde{\G}^{\m}_{\n\lambda} = - K^\m \!\left(\del_{(\n} L_{\lambda)} + L_{(\n} \pounds_K L_{\lambda)}\right) +\f{1}{2}\,H^{\m\rho} \left(\del_\n H_{\rho\lambda} + \del_\lambda H_{\rho\n} - \del_\rho H_{\n\lambda}\right) - H^{\m\rho} L_\lambda \widetilde{\mc{K}}_{\n\rho}\, .
\label{non_lcc}
\ee
Here $\widetilde{\mc{K}}_{\m\n} \equiv -\f{1}{2} \pounds_K H_{\m\n}$ is the extrinsic curvature for the PUL decomposition, which is purely spatial in nature i.e.~$K^\m \widetilde{\mc{K}}_{\m\n} = 0$. In particular, the Levi-Civita connection $\G^{\m}_{\n\lambda}$ can be expressed in terms of the PUL connection $\widetilde{\G}^\m_{\n\lambda}$ via
\be
\G^{\m}_{\n\lambda} = \f{1}{c^2} {\overset{\tiny{(-2)}}{\G}}{}^{\m}_{\n\lambda} + \widetilde{\G}^\m_{\n\lambda} + C^{\m}_{\n\lambda} + c^2 {\overset{\tiny{(2)}}{\G}}{}^{\m}_{\n\lambda}\,.
\label{lcc_decomp}
\ee
With $\widetilde{\G}^\m_{\n\lambda}$ given in eq.~\eqref{non_lcc}, the other terms appearing in eq.~\eqref{lcc_decomp} are\footnote{Here $``{\rm d}"$ denotes an exterior derivative: $({\rm d} L)_{\m\n} \equiv \del_\m L_\n - \del_\n L_\m$.}
\be
{\overset{\tiny{(-2)}}{\G}}{}^{\m}_{\n\lambda} = - K^\m \widetilde{\mc{K}}_{\n\lambda}\, , \quad C^\m_{\n\lambda} = H^{\m\rho} L_\lambda \widetilde{\mc{K}}_{\n\rho}\, , \quad {\overset{\tiny{(2)}}{\G}}{}^{\m}_{\n\lambda} = - L_{(\n} H^{\m\rho} ({\rm d} L)_{\lambda)\rho}\,.
\label{lcc_decomp_defs}
\ee
Note that no $c\to 0$ expansion has been performed in writing down eq.~\eqref{lcc_decomp} - it is simply a consequence of starting with the expression for the Levi-Civita connection in terms of the Lorentzian metric and its inverse and plugging in the PUL form eq.~\eqref{metric_pul}, followed by collecting terms with like powers of $c$. Using eq.~\eqref{lcc_decomp}, covariant derivatives involving the Levi-Civita connection can always be converted into covariant derivatives with respect to the PUL connection. This will be particularly useful in subsection \ref{carroll_spin_hydro}, when working out the equations of Carroll hydrodynamics starting from those of relativistic hydrodynamics, eq.~\eqref{hydro_eqs}, which involve the Levi-Civita covariant derivative.

Let us now turn to the hydrodynamic degrees of freedom and their PUL decomposition. The fluid four-velocity $u^\m$ can be decomposed as \cite{Armas:2023dcz}
\be
u^\m \equiv K^\m + c^2 \mathfrak{u}^\m\,,
\label{u_pul}
\ee
where $\mathfrak{u}^\m$ is arbitrary. The condition $u^\m u_\m \equiv u^\m u^\n g_{\m\n} = -c^2$ translates to 
\be
\label{norm_cond_1}
\mf{u}^\m \mf{u}^\n H_{\m\n} + 2 \mf{u}^\m L_\m - c^2 (\mf{u}^\m L_\m)^2 = 0.
\ee
Like $K^\m$, we will treat $\mf{u}^\m$ as a PUL variable, admitting an expansion in powers of $c^2$ when $c\to 0$ i.e.
\be
\begin{split}
&\mf{u}^\mu = \hat{\mf{u}}^\mu + \mc{O}(c^2) ,\\
&\mf{u}_\mu \equiv g_{\m\n} \mf{u}^\n = h_{\m\n} \hat{\mf{u}}^\nu + \mc{O}(c^2).
\end{split}
\label{PUL_carr_velo}
\ee
In the following, we will use the notation $\vec{\mf{u}}_\m \equiv h_{\m\n} \hat{\mf{u}}^\n$ to signify the purely spatial nature of this object. The condition eq.~\eqref{norm_cond_1} in the Carroll limit becomes
\be
\hat{\mf{u}}^\m \vec{\mf{u}}_\m + 2 \hat{\mf{u}}^\m \ell_\m = 0,
\label{carroll_norm_cond}
\ee
reducing the number of hydrodynamic degrees of freedom in $\hat{\mf{u}}^\m$ to three, as expected. For the spin degrees of freedom, we consider $\P_\m, \Pi_\m$ as PUL variables, and assume the following ansatz in the $c\to 0$ limit,
\be
\Phi_\m = \vec{\vp}_\m + \mc{O}(c^2)\, , \quad \Pi_\m = \vec{\pi}_\m + \mc{O}(c^2).
\label{PUL_spin_data}
\ee
The vector symbols on $\vec{\vp}_\m, \vec{\pi}_\m$ are once again to signify their purely spatial nature, which follows from the orthogonality conditions $u^\m\Phi_\m = u^\m\Pi_\m =0$ in the limit $c\to 0$, which give 
\be
k^\m\vec{\vp}_\m = 0\, , \quad  k^\m\vec{\pi}_\m = 0.
\label{ortho_conds}
\ee
Finally, for the energy density and pressure of the relativistic fluid appearing in the constitutive relation for the energy-momentum tensor eq.~\eqref{const_rels_1}, we will make the following ansatz,
\be
\e = \varepsilon + \mc{O}(c^2) \, , \quad P = p + \mc{O}(c^2) \, ,
\ee
in the limit $c\to 0$, where $\varepsilon, p$ respectively denote the energy density and pressure for the Carroll fluid.
 
Let us now examine the local Carroll boost transformation properties of the various Carrollian objects introduced above for the fluid degrees of freedom. The fluid four-velocity $u^\m$ is local Lorentz invariant, $\d_\Lambda u^\m = 0$, which, using the PUL decomposition eq.~\eqref{u_pul} and taking the $c\to 0$ limit implies that under a local Carroll boost $\d_\lambda \hat{\mf{u}}^\m = - \lambda^\m$, and thus $\d_\lambda \vec{\mf{u}}_\m = -\lambda_\m$. Further, we have $\d_\lambda \vec{\vp}_\m = \d_\lambda \vec{\pi}_\m = 0$. Importantly, the conditions eq.~\eqref{carroll_norm_cond} and \eqref{ortho_conds} are local Carroll boost invariant.

\subsection{Carroll hydrodynamics with a spin current}
\label{carroll_spin_hydro}
With the setup for describing the background Lorentzian geometry and the fluid degrees of freedom in terms of PUL variables in place, we can now proceed to work out the $c\to 0$ limit of the relativistic hydrodynamic equations \eqref{hydro_eqs}, to obtain the equations for Carroll hydrodynamics with a spin current. The strategy to do so would be to first express the relativistic energy-momentum tensor and the spin current, along with their conservation equations, in terms of PUL variables, followed by imposing the $c\to 0$ limit. Since the $c\to 0$ limit of PUL variables is already well understood, as discussed in detail in subsection \eqref{pul_setup}, obtaining the hydrodynamic equations in the Carroll limit becomes straight forward once the relativistic hydrodynamic equations have been expressed in terms of PUL variables. Let us first look at the energy-momentum sector. In terms of the PUL variables, the ideal relativistic fluid energy-momentum tensor eq.~\eqref{const_rels_1} can be written in a PUL decomposition as
\be
T^{\m}_{~\,\n} = {\overset{(0)}{T}}{\!}^{\m}_{~\,\n} + c^2 {\overset{(2)}{T}}{\!}^{\m}_{~\,\n} + c^4 {\overset{(4)}{T}}{\!}^{\m}_{~\,\n}\, ,
\label{EM_decomp_1}
\ee
with
\be
\begin{split}
{\overset{(0)}{T}}{\!}^{\m}_{~\,\n} &= (\e+P) K^\m (L_\n + H_{\n\lambda} \mf{u}^\lambda) + P\d^\m_{\n}\, , \\
{\overset{(2)}{T}}{\!}^{\m}_{~\,\n} &= (\e+P) \big[\mf{u}^\m(L_\n + H_{\n\lambda} \mf{u}^\lambda) - K^\m L_\n L_\lambda \mf{u}^\lambda\big]\, , \\
{\overset{(4)}{T}}{\!}^{\m}_{~\,\n} &= - (\e+P) \mf{u}^\m L_\n L_\lambda \mf{u}^\lambda.
\end{split}
\label{EM_decomp_2}
\ee
Next, we express the temporal and spatial projections of the energy-momentum conservation equations, i.e.~$K^\n \N_\m T^{\m}_{~\,\n} = 0$ and $H^{\rho\n} \N_\m T^\m_{~\,\n} = 0$, in a PUL decomposition, which yields
\begin{subequations}
\begin{align}
&K^\m \del_\m \e = (\e+P) \widetilde{\mc{K}} + \mc{O}(c^2),\label{pul_hydro_1}\\
&H^{\m\n} \del_\n P = - H^{\m\n} \Big[(\e+P)  (2K^\s \del_{[\s} L_{\n]} - \mf{u}^\s (\tilde{\mc{K}} H_{\s\n}  + \tilde{\mc{K}}_{\s\n})) \nonumber\\
&\qquad\qquad\qquad\qquad + K^\s \widetilde{\N}_\s((\e+P)H_{\n\lambda}\mf{u}^\lambda) \Big]+ \mc{O}(c^2), \label{pul_hydro_2}
\end{align}
\end{subequations}
where the $\mc{O}(c^2)$ terms have been omitted for brevity, as they will not contribute in the $c\to 0$ limit (see appendix \ref{full_terms} for their complete expressions). Also, $\tilde{\mc{K}} \equiv H^{\m\n} \tilde{\mc{K}}_{\m\n}$ is the trace of the extrinsic curvature in the PUL decomposition. 

In the limit $c\to 0$, eq.~\eqref{EM_decomp_1} with \eqref{EM_decomp_2} yields the energy-momentum tensor for an ideal Carroll fluid,
\be
\mc{T}^\m_{~~\n} = (\varepsilon + p) k^\m (\ell_\n + \vec{\mf{u}}_\n) + p \,\d^\m_{\n}.
\ee
Further, the $c\to 0$ limit of equations \eqref{pul_hydro_1} and \eqref{pul_hydro_2} can be straightforwardly computed to give
\begin{subequations}
\begin{align}
&k^\m \del_\m \varepsilon = (\varepsilon + p) \widehat{\mc{K}}\, ,\label{carr_hyd_1}\\
&h^{\m\n} \del_\n p = - (\varepsilon + p) (\xi^\m - \widehat{\mc{K}} h^{\m\n} \vec{\mf{u}}_\n) - h^{\m\n} k^\s \widehat{\N}_\s ((\varepsilon+p) \vec{\mf{u}}_\n),\label{carr_hyd_2}
\end{align}
\label{carr_hyd_pul}
\end{subequations}
where $\widehat{\mc{K}} \equiv h^{\m\n} \widehat{\mc{K}}_{\m\n}$ is the trace of the Carrollian extrinsic curvature, while $\xi^\m$ is given by $\xi^\m \equiv 2 h^{\m\n} k^\s \del_{[\s} \ell_{\n]} - h^{\m\n} \hat{\mf{u}}^\s \widehat{\mc{K}}_{\s\n}$, with the first term referred to as ``Carrollian acceleration.'' Eqs.~\eqref{carr_hyd_1} and \eqref{carr_hyd_2} are the equations of Carroll hydrodynamics for an ideal Carroll fluid carrying energy density $\varepsilon$ and pressure $p$, obtained previously in \cite{Armas:2023dcz}. Note that these must be further accompanied by an equation of state $p = p(\varepsilon)$, which will be a property of the specific Carroll fluid under consideration. 

We can now repeat a similar process to obtain the equations governing the spin current for the Carroll fluid, starting from the relativistic description. Consider first the spin density tensor eq.~\eqref{omega_decomp} written in the PUL decomposition,
\be
\O^{\m\n} = {\overset{(0)}{\O}}{\!}^{\m\n} + c^2 {\overset{(2)}{\O}}{\!}^{\m\n},
\label{omega_pul}
\ee
where\footnote{Recall that once we express quantities in terms of PUL variables, the spacetime indices $\m,\n,\ldots$ run over the coordinates $(t,x^i)$ instead of the numerals $(0,1,2,3)$. This fact is crucial to get the PUL decomposition eq.~\eqref{omega_pul} starting from eq.~\eqref{omega_decomp}. Due to this, a factor of $c$ appears in the denominator of the last term in eq.~\eqref{omega_decomp}, along with another factor of $c$ which follows from the PUL representation $\sqrt{-g} = c E$.}
\begin{subequations}
\begin{align}
{\overset{(0)}{\O}}{\!}^{\m\n} &= 2 K^{[\m} H^{\n]\rho} \P_\rho + \f{1}{E} {\varepsilon}^{\m\n\a\b} (\mf{u}^\rho H_{\rho\a}+L_\a) \Pi_\b \, ,  \label{pul_omega_1}\\
{\overset{(2)}{\O}}{\!}^{\m\n} &= 2 \mf{u}^{[\m} H^{\n]\rho} \P_\rho + 2\mf{u}^{[\m} K^{\n]} \mf{u}^\rho \P_\rho - \f{1}{E} {\varepsilon}^{\m\n\a\b} \mf{u}^\rho L_\rho L_\a \Pi_\b\,.\label{pul_omega_2}
\end{align}
\end{subequations}
Here $E \equiv \sqrt{{\rm det}(L_\m L_\n+H_{\m\n})}$ is the vierbein determinant. The ideal relativistic spin current, eq.~\eqref{const_rels_1}, then admits the following PUL decomposition,
\be
S^{\m\n\lambda} = {\overset{~(0)}{S}}{\!}^{\m\n\lambda} + c^2 \!\! {\overset{~(2)}{S}}{\!}^{\m\n\lambda} + c^4 \!\! {\overset{~(4)}{S}}{\!}^{\m\n\lambda},
\ee
where
\be
{\overset{~(0)}{S}}{\!}^{\m\n\lambda} = K^\m {\overset{(0)}{\O}}{\!}^{\n\lambda} , \quad {\overset{~(2)}{S}}{\!}^{\m\n\lambda} = \mf{u}^\m {\overset{(0)}{\O}}{\!}^{\n\lambda} + K^\m {\overset{(2)}{\O}}{\!}^{\n\lambda} , \quad {\overset{~(4)}{S}}{\!}^{\m\n\lambda} = \mf{u}^\m {\overset{(2)}{\O}}{\!}^{\n\lambda} ,
\ee
with ${\overset{(0)}{\O}}{\!}^{\n\lambda}, {\overset{(2)}{\O}}{\!}^{\n\lambda}$ given in eqs.~\eqref{pul_omega_1} and \eqref{pul_omega_2}, respectively. Using this, the spin current conservation equation \eqref{cons_eq_1} in PUL decomposition becomes
\be
K^\m \widetilde{\N}_\m {\overset{(0)}{\O}}{\!}^{\n\lambda} = \widetilde{\mc{K}} {\overset{(0)}{\O}}{\!}^{\n\lambda} - 2  \widetilde{\mc{K}}_{\m\s} \mf{u}^\s K^{[\n} {\overset{(0)}{\O}}{\!}^{\lambda]\m} + \mc{O}(c^2).
\ee
Computing the $c\to 0$ limit of the equation above is now straightforward, and yields
\be
k^\m \widehat{\N}_\m \mf{s}^{\n\lambda} = \widehat{\mc{K}}  \mf{s}^{\n\lambda} - 2  \widehat{\mc{K}}_{\m\s} \hat{\mf{u}}^\s k^{[\n} \mf{s}^{\lambda]\m} ,
\label{carroll_spin_eqn}
\ee
where 
\be
\mf{s}^{\m\n} \equiv \lim_{c\to 0} {\overset{(0)}{\O}}{\!}^{\m\n} = 2 k^{[\m} h^{\n]\rho} \vec{\varphi}_\rho + \f{1}{e} {\varepsilon}^{\m\n\a\b} (\vec{\mf{u}}_\a+\ell_\a) \vec{\pi}_\b 
\label{carroll_spin_density}
\ee
is the ``Carrollian spin density,'' with $e\equiv \sqrt{{\rm det}(\ell_\m \ell_\n+h_{\m\n})}$ being the vierbein determinant on the emergent Carrollian structure. Eqs.~\eqref{carroll_spin_eqn} and \eqref{carroll_spin_density}, giving the evolution equation for the Carrollian spin density for an ideal Carroll fluid endowed with spin, are some of the main results of this paper. One can further define the ideal ``Carrollian spin current'' as 
\be
\mc{S}^{\m\n\lambda} \equiv \lim_{c\to 0} S^{\m\n\lambda} = k^\m \mf{s}^{\n\lambda},
\label{carroll_spin_current}
\ee
in terms of which eq.~\eqref{carroll_spin_eqn} can be rewritten as 
\be
\widehat{\N}_\m \mc{S}^{\m\n\lambda} = - \widehat{\mc{K}}  \ell_\mu \mc{S}^{\m\n\lambda} - 2 \widehat{\mc{K}}_{\m\s} \hat{\mf{u}}^\s \mc{S}^{[\nu\lambda]\m} ,
\label{carroll_spin_eqn_2}
\ee
utilizing the compatibility of the Carrollian connection with $k^\m$, eq.~\eqref{connection_req}. Eq.~\eqref{carroll_spin_eqn_2} for the evolution of the ideal Carrollian spin current is the ultrarelativistic analog of the Lorentzian spin current evolution, eq.~\eqref{hydro_eqs}.\footnote{On taking the background to be the flat Carroll geometry, Carroll spin hydrodynamics also exhibits pseudogauge freedom, as discussed in appendix \ref{carr_pseudo}.} The fact that one ends up with a non-conservation equation is not surprising, as the Carroll compatible connection carries intrinsic torsion \cite{Gallegos:2021bzp, Hongo:2021ona, Gallegos:2022jow}. In fact, from eq.~\eqref{carr_comp_conn}, the torsion is given by
\be
\widehat{T}^{\m}_{~\,\n\lambda} \equiv \widehat{\G}^\m_{\n\lambda} - \widehat{\G}^\m_{\lambda\nu} =  h^{\m\rho} (\ell_\n \widehat{\mc{K}}_{\lambda\rho} - \ell_\lambda \widehat{\mc{K}}_{\nu\rho}),
\ee
which is completely antisymmetric in its lower two indices, and satisfies $\ell_\m \widehat{T}^{\m}_{~\,\n\lambda} = 0$. The extrinsic curvature and its trace can be expressed in terms of the torsion via
\be
\widehat{\mc{K}}_{\m\n} = h_{\n\lambda} k^\rho \widehat{T}^{\lambda}_{~\,\m\rho}\, , \quad \widehat{\mc{K}} = k^\n \widehat{T}^\m_{~\,\m\n}.
\ee
Thus, the ideal Carrollian spin current evolution equation \eqref{carroll_spin_eqn_2} can be recast in terms of the background torsion as
\be
\widehat{\N}_\m \mc{S}^{\m\n\lambda} = - k^\g \widehat{T}^\a_{~\,\b\g} \left(\d^\b_\a \ell_\m \mc{S}^{\m\n\lambda} + 2 h_{\m\a} \hat{\mf{u}}^\b \mc{S}^{[\n\lambda]\m}\right).
\label{carroll_spin_eqn_3}
\ee
The above form of the Carroll spin current evolution equation makes it explicit that non-conservation of the spin current is sourced directly by the non-vanishing background torsion.

An important aspect relevant for later discussion is the conformal limit of the hydrodynamic construction. For the relativistic theory in the conformal limit, the energy-momentum tensor must be traceless, $T^\m_{~\m} = 0$, which implies the equation of state $\e = 3P$ using the constitutive relations eq.~\eqref{const_rels_1}. In the Carroll limit, the equation of state for an ideal conformal Carroll fluid then simply becomes $\varepsilon = 3p$. Further, following  \cite{Shapiro:2001rz}, in the conformal limit for a relativistic fluid with a spin current, one must have $\N_\m S_\n^{~\,\n\m} = 0$, which for the ideal spin current constitutive relation in eq.~\eqref{const_rels_1}, combined with the decomposition eq.~\eqref{omega_decomp}, yields $\N_\m \P^\m = 0$. One can express this condition in PUL decomposition as\footnote{There are no $\mc{O}(c^2)$ terms in the PUL decomposition of the conformal constraint eq.~\eqref{PUL_CC}.}
\be
H^{\m\n} \widetilde{\N}_\m \P_\n + (K^\m \del_\m - \widetilde{\mc{K}}) \mf{u}^\n \P_\n + K^\m H^{\n\lambda} \Phi_{\lambda} (\del_\m L_\n - \del_\n L_\m) = 0.
\label{PUL_CC}
\ee
On taking the limit $c\to 0$, this condition becomes
\be
h^{\m\n}\widehat{\N}_\m \vec{\varphi}_\n + (k^\m \del_\m-\widehat{\mc{K}}) \hat{\mf{u}}^\nu \vec{\varphi}_\nu + k^\m h^{\n\lambda} \vec{\varphi}_\lambda (\del_\m \ell_\n - \del_\n \ell_\m) = 0 ,
\label{spin_conf_const}
\ee
which can be written more succinctly as 
\be
\widehat{\N}_\m (h^{\m\n} \vec{\varphi}_\n) + (k^\m \del_\m-\widehat{\mc{K}}) \hat{\mf{u}}^\nu \vec{\varphi}_\nu = 0.
\label{spin_conf_const_2}
\ee
An ideal Carroll fluid with a spin current must satisfy the above in the conformal limit.

Before we conclude this section, a word about the local Carroll boost transformation properties of the Carrollian spin density and the spin current. Using eqs.~\eqref{lcb_trans} and \eqref{lcb_trans_2}, along with $\d_\lambda \vec{\mf{u}}_\m = -\lambda_\m$ while $\d_\lambda \vec{\vp}_\m = \d_\lambda \vec{\pi}_\m = 0$, it is straight forward to check that the Carrollian spin density $\mf{s}^{\m\n}$, eq.~\eqref{carroll_spin_density}, as well as the spin current $\mc{S}^{\m\n\lambda}$, eq.~\eqref{carroll_spin_current}, are local Carroll boost invariant, as the case should be for physically relevant entities. In particular, the combination $(\ell_\m+\vec{\mf{u}}_\m)$ is invariant under local Carroll boosts. For a detailed discussion of the local Carroll boost invariance of the Carroll hydrodynamic equations \eqref{carr_hyd_1}, \eqref{carr_hyd_2}, and \eqref{carroll_spin_eqn}, along with the invariance of the conformal constraint eq.~\eqref{spin_conf_const}, please refer to appendix \ref{symmetries}. 

\section{Mapping to Bjorken and Gubser flow with spin}
\label{boost_inv_models}
Ultrarelativistic heavy-ion collisions provide an interesting arena to explore nature at extreme energy scales. The dynamics that ensues from these collisions is very complex, and can be modeled as going through several intermediate phases before the eventual appearance of free streaming particles that are observed by the detectors. The highly non-equilibrium phase immediately after the collision is succeeded by the QGP phase, once local thermal equilibrium is achieved, and relativistic hydrodynamics becomes a good description of the ongoing physics. To gain an analytic understanding of QGP dynamics, simplified hydrodynamic models have been proposed in the past, by imposing various phenomenological symmetries on the QGP flow. The most prominent amongst these are the Bjorken flow \cite{PhysRevD.27.140} and Gubser flow \cite{PhysRevD.82.085027, Gubser:2010ui} models. In Bjorken flow, one assumes that all the interesting aspects of QGP hydrodynamics happen along the longitudinal beam axis, while any transverse dynamics can be ignored. Thus, in Bjorken flow, one has complete translation and rotation invariance in the plane transverse to the beam axis. The assumption of exact translation invariance is relaxed in Gubser flow, assuming the fluid to be conformal, by demanding invariance only under combinations of translations and special conformal transformations in the transverse plane,\footnote{More precisely, one demands invariance up to a conformal factor.} along with rotation invariance about the beam axis. This allows for the flow to acquire a nontrivial radial profile as well. 

\begin{figure}[t!]
\centering
\includegraphics[scale=1.15]{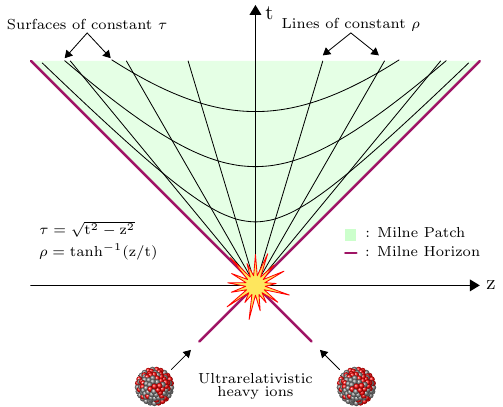}
\caption{Depiction of the heavy-ion collision process. The collision happens at time $t=0$ at the origin. Without loss of generality, one can align the $z$-axis along the beam direction, while the $(x,y)$-axes form the transverse plane. The Milne patch of the Minkowski spacetime, covered by the proper time, rapidity coordinates $(\tau, \rho)$, serves as the forward lightcone for the collision event.}
\label{fig:lightcone}
\end{figure}

Apart from these assumptions, both Bjorken and Gubser flow presume invariance of the flow under boosts along the beam axis, which is reasonably well motivated from the observed data. Working in Milne coordinates, fig.~\ref{fig:lightcone}, with $\tau=\sqrt{t^2 -z^2}$ being the proper time and $\rho = \tanh^{-1}(z/t)$ being the rapidity, where we have aligned the Cartesian $z$-axis along the beam while $(x, y)$ axes form the transverse plane, the assumption of invariance under boosts along the beam axis translates into independence of the flow from the rapidity $\rho$. With these assumptions in place, one can completely determine the fluid four-velocity profile based entirely on symmetries, which can then be used as an input in the relativistic hydrodynamic equations to determine the spacetime evolution of the QGP energy density and pressure. Though Bjorken and Gubser flow are simplified models, they serve a very important purpose by providing an intuitive understanding of the actual complex dynamics of the QGP in heavy-ion collisions.

Interestingly, it was realized in \cite{Bagchi:2023ysc, Bagchi:2023rwd} that the equations for Bjorken and Gubser flow can be obtained from that of a Carroll fluid by making suitable choices for the background Carroll geometry. In other words, Bjorken and Gubser flow serve as potential real world examples of Carroll hydrodynamics. The underlying reason why this mapping at the level of equations exists between the two is because of the large number of symmetry assumptions inherent in Bjorken and Gubser flow, which effectively render the relativistic fluid equations into a form equivalent to performing a 3+1-like split, characteristic of the Carroll hydrodynamic setup. In the present section, we present the maps that take one from the equations of Carroll hydrodynamics obtained using the PUL parametrization and the $c\to 0$ limit, eqs.~\eqref{carr_hyd_pul} and \eqref{carroll_spin_eqn_2}, to Bjorken and Gubser flow, now with the inclusion of a spin current. The analogous discussion in terms of the PR parametrization is given in appendix \ref{pr_gauge}. 

\subsection{Bjorken flow with a spin current}
\label{bjorken}
Let us first consider Bjorken flow with a spin current. It is convenient to set this up in terms of the Milne coordinates $(\tau, \rho, x, y)$, fig.~\ref{fig:lightcone}, in terms of which the background metric takes the form
\be
ds^2 = -d\tau^2 + \tau^2 d\rho^2 + dx^2 + dy^2.
\label{milne_metric}
\ee
Demanding the flow profile i.e.~the fluid four-velocity $u^\m$, normalized such that $u^\m u_\m = -1$, to respect translation and rotation invariance in the traverse plane, as well as invariance under boosts along the beam axis, uniquely fixes it to the form $u^\m = (1,0,0,0)$ in Milne coordinates i.e.~the fluid appears static. This follows from the fact that the invariance of $u^\m$ under a spacetime transformation $x^\m \rightarrow x^\mu + \chi^\m(x)$ i.e.~with the generator $\chi\equiv \chi^\m \del_\m$ amounts to the requirement that $\pounds_\chi u^\m = 0$. For Bjorken flow, the symmetry generators under which $u^\m$ must be invariant are essentially the isometries of the background Milne metric eq.~\eqref{milne_metric} i.e.~$\del_x, \del_y, x\del_y - y\del_x$ and $\del_\rho$,\footnote{The Milne background is Minkowski spacetime covered by an expanding coordinate chart. It has ten isometries, which together generate the Poincar\'{e} algebra. However, the additional six isometry generators not imposed as symmetries on Bjorken flow do not preserve the constant-$\tau$ hypersurfaces, and mix $\tau$ with $\rho, x, y$. In other words, they do not maintain the 3+1-like split of the background geometry, where the fluid lives on constant-$\tau$ spatial hypersurfaces and evolves along the orthogonal direction $\del_\t$. Imposing only the background isometries that preserve the 3+1-like split as symmetries on Bjorken flow plays a crucial role in making it appear effectively Carrollian.} resulting in the static profile $u^\m = (1,0,0,0)$. Substituting this in the constitutive relation eq.~\eqref{const_rels_1} for the energy-momentum tensor of an ideal relativistic fluid and computing the hydrodynamic equations $\N_\m T^{\m\n} = 0$ yields
\be
\f{d\e}{d\t} = - \f{\e+P}{\t}.
\label{Bjorken_eq_1}
\ee
More specifically, the above follows from the $\nu = \tau$ component of $\N_\m T^{\m\n} = 0$, while the spatial components yield $\del_i P = 0$, $i = (\rho,x,y)$, implying that the pressure is independent of the rapidity and transverse coordinates, which are the underlying assumptions for Bjorken flow. Given that energy density and pressure are related via an equation of state, $P = P(\epsilon)$, one can conclude that the energy density is also independent of the rapidity and transverse coordinates, and is thus only a function of the proper time $\t$, which was utilized by replacing the partial derivative with an ordinary one in eq.~\eqref{Bjorken_eq_1}. With the knowledge of the equation of state and the initial conditions at some proper time $\tau_0$, eq.~\eqref{Bjorken_eq_1} can be solved to compute the time evolution of the energy density of the QGP. 

Let us now derive the equations governing the spin current using Bjorken's phenomenological symmetries. As discussed in subsection \ref{rel_hydro}, the dynamical degrees of freedom in the spin current are encoded in the spacelike vector fields $(\Phi^\m, \Pi^\m)$, eq.~\eqref{omega_decomp}. By demanding that they satisfy the symmetries of Bjorken flow, one can uniquely fix them to the form
\be
\Phi^\m = f(\tau) \left(0,\f{1}{\t},0,0\right), \quad \Pi^\m = g(\tau) \left(0,\f{1}{\t},0,0\right),
\label{phi_pi_bjorken}
\ee
where $f(\tau)$ and $g(\tau)$ are arbitrary functions of proper time. Note that the vector $(0,\tau^{-1},0,0)$ is the unique spacelike unit-vector that respects Bjorken's symmetry assumptions i.e. translation and rotation invariance in the transverse plane along with independence from rapidity. Thus, for Bjorken flow, the spin density $\O^{\m\n}$, and consequently the spin current $S^{\m\n\lambda}$, have only two degrees of freedom instead of six, denoted by $(f(\tau), g(\tau))$. Computing the equations for the conservation of spin current, $\N_\m S^{\m\n\lambda} = 0$, with the constitutive relation for the spin current in eqs.~\eqref{const_rels_1} and \eqref{omega_decomp}, combined with eq.~\eqref{phi_pi_bjorken} as well as the fluid four-velocity profile $u^\m = (1,0,0,0)$ for Bjorken flow, yields 
\be
\tau \f{df(\tau)}{d\tau} + f(\tau) = 0\, , \quad \tau \f{dg(\tau)}{d\tau} + g(\tau) = 0\, ,
\label{bj_spin_eqs_1}
\ee
or equivalently
\be
\f{d}{d\tau}\big(\t f(\t)\big) = 0\, , \quad \f{d}{d\tau}\big(\t g(\t)\big) = 0\,.
\label{bj_spin_eqs_2}
\ee
These can be easily solved to get $f(\tau) = f_0/\tau$ and $g(\tau) = g_0/\tau$, with $(f_0, g_0)$ constant.

\subsection{Gubser flow with a spin current}
\label{gubser}
Next, let us consider Gubser flow with a spin current. As mentioned earlier, Gubser flow retains the requirements of boost invariance along, and rotation invariance about, the beam axis, but relaxes the requirement of exact translation invariance in the transverse plane to conformal invariance under combinations of translations and special conformal transformations. More specifically, the desired (conformal) symmetry generators in the transverse plane for Gubser flow are \cite{PhysRevD.82.085027}
\be
\chi_1 \equiv \del_x + q^2 (2xx^\m \del_\m - x^\m x_\m \del_x), \quad \chi_2 \equiv \del_y + q^2 (2yx^\m \del_\m - x^\m x_\m \del_y),
\label{x1x2}
\ee
where $q$ is a tunable parameter carrying dimensions of inverse length, with $q=0$ leading to the generators $\del_x, \del_y$ of Bjorken flow. Thus, $(\chi_1, \chi_2)$ are linear combinations of the translation generators with those of special conformal transformations, $2b^\n x_\n x^\m \del_\m - x^\m x_\m b^\n \del_\n$, with $b^\n = (\d^\n_x, \d^\n_y)$, on the transverse plane. They constitute part of the conformal isometries for the background Milne metric, as they satisfy
\be
\pounds_{\chi_a} g_{\m\n} = \f{1}{2} (\N_\lambda \chi^\lambda_a) g_{\m\n}, ~\text{with} ~~ a = (1,2).
\ee
In fact, together with the generator of rotations about the beam axis, $\chi_{\rm rot} \equiv x\del_y - y\del_x$, they form an $\mathfrak{so}(3)_q$ subalgebra of the four-dimensional conformal algebra $\mathfrak{so}(4,2)$. To wit
\be
[\chi_1, \chi_2] = - 4q^2 \chi_{\rm rot}\, , \quad [\chi_1, \chi_{\rm rot}] = \chi_2\, , \quad \quad [\chi_2, \chi_{\rm rot}] = - \chi_1\, .
\ee
Further, the $\mathfrak{so}(3)_q$ generators commute with the boost generator along the beam axis, $\chi_{\rm boost} \equiv z\del_t+t\del_z $ ($\equiv \del_\rho$ in Milne coordinates), which forms an $\mathfrak{so}(1,1)$ subalgebra of $\mathfrak{so}(4,2)$. In Milne coordinates, demanding invariance of the fluid four-velocity under rotations about the beam axis and boosts along the beam axis, along with conformal invariance in the transverse plane under $(\chi_1, \chi_2)$ fixes $u^\m$ to be a function of $(\tau, r)$, with $r=\sqrt{x^2+y^2}$ being the radial coordinate in the transverse plane. This can then be inserted into the constitutive relation for the energy-momentum tensor, along with the conformal equation of state $\e =3P$, to derive the hydrodynamic equations for Gubser flow, which now give a set of coupled partial differential equations determining the evolution of energy density as a function of $(\tau, r)$ \cite{PhysRevD.82.085027}. 

Gubser flow can also be expressed, perhaps more elegantly, on the global dS$_3\times\mathbb{R}$ background \cite{Gubser:2010ui}, which is the formulation we will focus on in the present work. One can arrive at the global dS$_3\times\mathbb{R}$ background by first Weyl rescaling the Milne metric eq.~\eqref{milne_metric} via $ds^2 \to ds^2/\tau^2$, followed by a coordinate transformation from the $(\tau, r)$ coordinates to the $(\varsigma, \psi)$ coordinates via
\be
\sinh \varsigma = - \f{1-q^2(\t^2-r^2)}{2q\tau}\, , \quad \tan\psi = \f{2qr}{1+q^2(\t^2-r^2)}\,.
\ee 
As a consequence of the above transformations, the background metric becomes that of the global dS$_3\times\mathbb{R}$ spacetime,
\be
ds^2 = - d\varsigma^2 + \cosh^2\varsigma \, (d\psi^2 + \sin^2 \psi \, d\phi^2) + d\rho^2\, ,
\label{global_dSR}
\ee
with $\phi$ being the angular coordinate about the beam axis.\footnote{$\phi \equiv \tan^{-1}(y/x)$ in terms of Cartesian coordinates in the transverse plane.} The primary advantage of working in the global dS$_3\times\mathbb{R}$ formulation for Gubser flow is that the $\mathfrak{so}(3)_q$ conformal symmetry generators for the Milne background now form exact isometries of the background metric eq.~\eqref{global_dSR}, associated with the rotational symmetry on the two-sphere parametrized by the coordinates $(\psi, \phi)$. The shifts in rapidity (i.e.~boosts along the beam axis) are still an exact isometry, as they were in the Milne background eq.~\eqref{milne_metric}. One can now demand the fluid four-velocity profile to be invariant under these background isometries, which leads to $u^\m = (1,0,0,0)$ i.e.~Gubser flow is static on the global dS$_3\times\mathbb{R}$ background, which is another advantage offered by this formulation.\footnote{The global dS$_3\times\mathbb{R}$ background has seven isometry generators. The additional three generators that are not imposed as symmetries on Gubser flow mix $\varsigma$ with the coordinates on the two-sphere $(\psi, \phi)$ i.e. they do not preserve the 3+1-like split of the background, where the fluid lives on constant-$\varsigma$ spatial slices and evolves along the orthogonal $\del_\varsigma$ direction. Like Bjorken flow, imposing only those background isometries as symmetries of the Gubser flow that respect the 3+1-like split plays a crucial role in its mapping to a Carroll fluid.} Inserting this velocity profile into the constitutive relation for the energy-momentum tensor eq.~\eqref{const_rels_1}, along with the conformal equation of state $\e=3P$, yields the following hydrodynamic equation,
\be
\f{d\e}{d\varsigma} = - \f{8\e}{3} \tanh\varsigma\,.
\label{Gubser_eq_1}
\ee
More specifically, the above equation follows from the $\nu = \varsigma$ component of the hydrodynamic equations $\N_\m T^{\m\n} = 0$, while the other components give $\del_i \e= 0, i = (\rho, \psi, \phi)$, implying that the energy density is independent of the spacelike coordinates, which was used in converting the partial derivative in eq.~\eqref{Gubser_eq_1} to an ordinary one. Eq.~\eqref{Gubser_eq_1} can be easily solved to obtain the evolution of the QGP energy density, and gives $\e = \e_0(\sech\varsigma)^{8/3}$. 

Let us now work out the equations for the spin current for Gubser flow on the global dS$_3\times\mathbb{R}$ background. The unique spacelike unit-vector compatible with the phenomenological symmetries of Gubser flow is $(0,0,0,1)$. Therefore, the most general form taken by the fields $(\Phi^\m, \Pi^\m)$ on the global dS$_3\times\mathbb{R}$ background compatible with Gubser symmetries is
\be
\Phi^\m = v(\varsigma) (0,0,0,1)\, , \quad \Pi^\m = w(\varsigma) (0,0,0,1)\, ,
\label{phipidS}
\ee
where $v(\varsigma)$ and $w(\varsigma)$ are arbitrary functions of $\varsigma$. Once again, the spin density and the spin current comprise only two degrees of freedom, $v(\varsigma)$ and $w(\varsigma)$. Inserting eq.~\eqref{phipidS} along with $u^\m = (1,0,0,0)$ into the constitutive relation for the spin current eqs.~\eqref{const_rels_1} and \eqref{omega_decomp}, and computing the spin current conservation equation $\N_\m S^{\m\n\lambda} = 0$ yields
\be
\f{dv(\varsigma)}{d\varsigma} + 2\tanh\varsigma \, v(\varsigma) = 0\, , \quad \f{dw(\varsigma)}{d\varsigma} + 2\tanh\varsigma \,w(\varsigma) = 0\, .
\label{gubser_spin_dofs}
\ee
They admit the simple solutions $v(\varsigma) = v_0 \sech^2 \varsigma$, $w(\varsigma) = w_0 \sech^2 \varsigma$, with $(v_0, w_0)$ constant. Further, the conformal constraint $\N_\m S_\n^{~\,\n\m} = 0$ is satisfied identically and does not impose any additional constraint on $v(\varsigma)$ and $w(\varsigma)$.

\subsection{From Carroll hydrodynamics to Bjorken and Gubser flow with spin}
\label{the_maps}
Let us now state the maps that take one from the equations of ideal Carroll hydrodynamics with spin, eqs.~\eqref{carr_hyd_pul} and \eqref{carroll_spin_eqn}, to the corresponding equations for Bjorken and Gubser flow with spin. 

Consider the following choice for the geometric and fluid data of the ideal Carroll fluid in  eq.~\eqref{carr_hyd_pul}, where spacetime indices $\m,\n,\ldots$ on the Carroll manifold take  values $(\t,\rho,x,y)$,
\be
k^\m \del_\m = \del_\tau\, , \qquad~ h_{\m\n} dx^\m dx^\n = \tau^2 d\rho^2 + dx^2 + dy^2\, , \quad (\ell_\m + \vec{\mathfrak{u}}_\m) dx^\m = -d\tau\, .
\label{bj_data}
\ee
The choice above is adapted to the 3+1-like split inherent in the Bjorken flow setup. In particular, the kernel $k^\m$ is aligned along the fluid four-velocity in Bjorken flow, while the spatial metric $h_{\m\n}$ on the Carroll manifold is same as the induced metric on constant-$\tau$ hypersurfaces of the Milne background. Further, the local Carroll boost invariant combination $(\ell_\m + \vec{\mathfrak{u}}_\m)dx^\m$ is chosen to measure the passage of time $-d\tau$, with the minus sign fixed by the requirement $k^\m \ell_\m = -1$. Note that we do not individually fix the components $\ell_i$ and $\vec{\mathfrak{u}}_i$,\footnote{The choices in eq.~\eqref{bj_data} fix $\ell_\t = -1$, as $\vec{\mathfrak{u}}_\t = 0$ due to its purely spatial nature.} but only their local Carroll boost invariant combination, $\ell_i + \vec{\mathfrak{u}}_i = 0$, to maintain full generality. Next, by using the orthogonality and completeness relations satisfied by the co-metric $h^{\m\n}$, eq.~\eqref{completeness_2}, we can fix it to the form\footnote{This can be obtained as follows. Using orthogonality, $h^{\m\n} \ell_\n = 0$, one can solve for $(h^{\t\t}, h^{\t i})$ in terms of $h^{ij}$ viz.~ $h^{\t\t} = h^{ij}\ell_i\ell_j, h^{\t i} = h^{ij} \ell_j$. Next, using orthogonality, $h^{\m\n} h_{\n\lambda} - k^\m \ell_\lambda = \d^\m_\lambda$, one can show that $h^{ij} h_{jk} = \d^i_k$. These results fix $h^{\m\n}$ to the form eq.~\eqref{bj_hinv}. Note that unlike the data fixed in eq.~\eqref{bj_data}, $h^{\m\n}$ is not local Carroll boost invariant, but the eqs.~\eqref{carr_hyd_pul} and \eqref{carroll_spin_eqn} are, as elaborated upon in appendix \ref{symmetries}.}
\be
h^{\m\n} \del_\m \del_\n = (h^{ij}\ell_i \ell_j) \del_\t^2 + (2 h^{ij} \ell_j) \del_\t \del_i + h^{ij} \del_i \del_j\, , \quad h^{ij} \del_i \del_j = \f{1}{\t^2} \del_\rho^2 + \del_x^2 + \del_y^2.
\label{bj_hinv}
\ee
Specializing the ideal Carroll fluid in eq.~\eqref{carr_hyd_pul} to the data chosen in eqs.~\eqref{bj_data} and \eqref{bj_hinv}, one gets the equations 
\be
\del_\t \varepsilon = - \f{\varepsilon + p}{\t}\, , \quad \del_i p = 0.
\label{bj_carroll}
\ee
These equations are immediately recognizable as depicting Bjorken flow, eq.~\eqref{Bjorken_eq_1}. In particular, the second equation, $\del_i p = 0$, captures the phenomenological assumptions of Bjorken flow viz.~independence from the rapidity $\rho$ as well as translations and rotations in the transverse plane, as the pressure $p$, and consequently the energy density $\varepsilon$, depend only upon $\t$. Additionally, if one chooses the fields $(\vec{\varphi}_\m, \vec{\pi}_\m)$ appearing in the spin density eq.~\eqref{carroll_spin_density} for the Carroll fluid to be of the form
\be
\vec{\varphi}_\m dx^\m = \t \hat{f}(\t) \, d\rho\, , \quad \vec{\pi}_\m dx^\m = \t \hat{g}(\t)\, d\rho\, ,
\label{pul_bj_spin_data}
\ee
along with the data eqs.~\eqref{bj_data} and \eqref{bj_hinv}, then eq.~\eqref{carroll_spin_eqn} for the Carrollian spin current implies
\be
\tau \f{d\hat{f}(\tau)}{d\tau} + \hat{f}(\tau) = 0\, , \quad \tau \f{d\hat{g}(\tau)}{d\tau} + \hat{g}(\tau) = 0\, .
\ee
The above equations are of the same form as the eqs.~\eqref{bj_spin_eqs_1} for the spin degrees of freedom in Bjorken flow. In other words, the choice of geometric and fluid data in eqs.~\eqref{bj_data}, \eqref{bj_hinv} and \eqref{pul_bj_spin_data} maps the equations for an ideal Carroll fluid endowed with a spin current to that of Bjorken flow with spin. This extends the mapping first uncovered in \cite{Bagchi:2023ysc} between Carroll hydrodynamics and Bjorken flow to include a spin current as well.

Let us discuss next the mapping of Carroll hydrodynamics to Gubser flow on the global dS$_3\times\mathbb{R}$ background, with the inclusion of spin. Consider the following choice for the geometric and fluid data for the ideal Carroll fluid endowed with spin in eqs.~\eqref{carr_hyd_pul} and \eqref{carroll_spin_eqn}, with the spacetime indices $\m,\n,\ldots$ running over $(\varsigma,\psi,\phi,\rho)$ on the Carroll manifold,
\begin{align}
&k^\m \del_\m = \del_\varsigma \, , \qquad~ h_{\m\n} dx^\m dx^\n = \cosh^2\varsigma\, (d\psi^2 + \sin^2\psi \, d\phi^2) + d\rho^2\, , \quad (\ell_\m + \vec{\mathfrak{u}}_\m) dx^\m = -d\varsigma\, ,\nonumber\\
&h^{\m\n} \del_\m \del_\n = (h^{ij}\ell_i \ell_j) \del_\varsigma^2 + (2 h^{ij} \ell_j) \del_\varsigma \del_i + h^{ij} \del_i \del_j\, , \quad h^{ij} \del_i \del_j =\sech^2\varsigma\, (\del_\psi^2 + \rm{cosec}^2 \psi \, \del_\phi^2) + \del_\rho^2 ,\nonumber\\
&\vec{\varphi}_\m dx^\m = \hat{v}(\varsigma) d\rho \, , \quad \vec{\pi}_\m dx^\m = \hat{w}(\varsigma) d\rho\,. \label{gb_data}
\end{align}
Once again, similar to the discussion for mapping to Bjorken flow, we have aligned the kernel $k^\m$ with the fluid four-velocity for Gubser flow on the global dS$_3\times\mathbb{R}$ background, while the spatial metric $h_{\m\n}$ on the Carroll manifold is same as the induced metric on constant-$\varsigma$ hypersurfaces of the global dS$_3\times\mathbb{R}$ background. The local Carroll boost invariant combination $(\ell_\m + \vec{\mathfrak{u}}_\m)dx^\m$ measures the passage of time $-d\varsigma$, while the co-metric $h^{\m\n}$ is determined using the orthogonality and completeness relations in eq.~\eqref{completeness_2}. The choices in eq.~\eqref{gb_data} when imposed on eqs.~\eqref{carr_hyd_pul} and \eqref{carroll_spin_eqn} with the conformal equation of state $\varepsilon = 3p$ yield
\be
\del_\varsigma \varepsilon = - \f{8\varepsilon}{3} \tanh \varsigma \, , \quad \del_i \varepsilon = 0\, ,
\label{gb_fluid_eqs}
\ee
along with
\be
\f{d\hat{v}(\varsigma)}{d\varsigma} + 2\tanh\varsigma \, \hat{v}(\varsigma) = 0\, , \quad \f{d\hat{w}(\varsigma)}{d\varsigma} + 2\tanh\varsigma \,\hat{w}(\varsigma) = 0\, .
\label{gb_spin_eqs}
\ee
Eqs.~\eqref{gb_fluid_eqs} are the equations for Gubser flow on the global dS$_3\times\mathbb{R}$ background. In particular, the second equation, $\del_i \varepsilon = 0$, asserts the independence of the energy density from the rapidity $\rho$ as well as the angles on the two-sphere $(\psi,\phi)$, which are the phenomenological symmetries imposed in Gubser flow, while the first equation in eq.~\eqref{gb_fluid_eqs} is the dynamical equation \eqref{Gubser_eq_1}, obtained from a Carrollian perspective by making appropriate geometric choices, eq.~\eqref{gb_data}. Further, the eqs.~\eqref{gb_spin_eqs} have the same form as the equations governing the spin degrees of freedom in Gubser flow, eqs.~\eqref{gubser_spin_dofs}. Note that the conformal constraint is identically satisfied for the data eq.~\eqref{spin_conf_const}, and does not impose any additional constraint on the Carrollian spin degrees of freedom $(\hat{v}, \hat{w})$, which is also the case for Gubser flow. The above results extend the mapping found in \cite{Bagchi:2023rwd} between Carroll hydrodynamics and Gubser flow, now with the inclusion of a spin current.

\section{Discussion and outlook}
\label{discussion}
In this paper, we have initiated the study of spin currents in Carroll hydrodynamics. Obtained by imposing the $c\to 0$ limit on the equation governing the spin current in relativistic hydrodynamics, either in the PUL or the PR parametrization, eqs.~\eqref {carroll_spin_eqn} and \eqref{spin_eqs_PR} respectively, these equations express the evolution of a local Carrollian spin density on the Carroll manifold. Further, we have expanded the maps discovered in \cite{Bagchi:2023ysc, Bagchi:2023rwd} between Carroll hydrodynamics and boost-invariant models for the spacetime evolution of QGP in heavy-ion collisions, namely Bjorken and Gubser flow, such that the equation for the spin current of the Carroll fluid maps to the corresponding one in these models by appropriately choosing the spin degrees of freedom.

There are a plethora of directions to explore further. For instance, our focus in the present work has been limited to ideal fluids, with no derivative terms in the constitutive relations for the energy-momentum tensor or the spin current, which can lead to dissipation. It would be interesting to expand the discussion with the inclusion of derivative corrections in the constitutive relations, especially for the spin current, where treatments based on different approaches to relativistic spin hydrodynamics advocate different possible terms that can appear at first order in derivatives. These approaches are for instance either based on kinetic theory and Boltzmann transport \cite{Bhadury:2020puc, Bhadury:2020cop, Weickgenannt:2022qvh, Drogosz:2024gzv}, or based on allowing all possible first order in derivative terms constrained by positivity of entropy production \cite{Hattori:2019lfp, Biswas:2023qsw, Becattini:2023ouz, Daher:2025pfq, Becattini:2025oyi, Fang:2025aig}, or based on the quantum statistical operator formulation \cite{Becattini:2012pp, Hu:2021lnx}, or based on sourcing the currents by coupling to a torsionful background geometry \cite{Gallegos:2021bzp, Hongo:2021ona, Gallegos:2022jow} etc.~The form of the constitutive relations of course also depends upon the choice of the hydrodynamic frame as well as the pseudogauge. Thus, at the technical level, to include derivative corrections one has to start by choosing a first order relativistic spin hydrodynamic framework and its associated constitutive relations, followed by applying the PUL/Carrollian limiting procedure discussed in this work to them. The resulting Carrollian theory would then contain, in addition to the ideal terms considered here, derivative corrections to the energy-momentum tensor and to the spin current. In particular, first order corrections to the spin current may contain spin diffusion or relaxation effects, as well as couplings to gradients of the hydrodynamic variables, vorticity, acceleration etc. These terms would modify the ideal energy-momentum and spin current equations for the Carroll fluid obtained in the present analysis. The ideal results presented here thus provide the starting point for a systematic extension incorporating derivative corrections, which we plan to pursue in the near future. It would be worthwhile to see how the mapping between Carroll fluids and Bjorken/Gubser flow with spin works out after the inclusion of derivative corrections in the constitutive relations.

Another useful direction to pursue would be to construct ideal Carroll hydrodynamics with a spin current using an equilibrium generating functional approach, along the lines of the discussion in \cite{Armas:2023dcz}. It was observed in \cite{Armas:2023dcz} that the generating functional approach, after properly accounting for the Goldstone bosons of spontaneously broken Carroll boost invariance, allows for two distinct classes of Carroll fluids: one which can be obtained by imposing the $c\to 0$ limit on the equations of relativistic hydrodynamics, which has been the methodology followed in the present work to construct Carroll hydrodynamics with spin, and another which satisfy an equation of state of the form $\varepsilon + p = 0$. Constructing the generating functional for Carroll hydrodynamics, now including the spin degrees of freedom, where the spin current can be thought of as sourced by the spin connection on the Carroll manifold, or equivalently by the torsion inherent in the Carrollian connection eq.~\eqref{carr_comp_conn}, it would be interesting to work out the properties of this second class of Carroll fluids endowed with spin.

A more microscopic perspective on spin currents on Carroll manifolds is also worth pondering over, as opposed to the macroscopic viewpoint offered by Carroll hydrodynamics. Carroll fermions \cite{Hao:2022xhq, Banerjee:2022ocj, Bagchi:2022eui, Bergshoeff:2023vfd, Ekiz:2025hdn, Grumiller:2025rtm} have been the subject of discussion in recent literature, partly because of their potential connections to condensed matter systems with flat bands. One may couple Carroll fermion theories to background Carroll gravity, and obtain the expressions and analyze the structure of the spin currents this gives rise to. The construction might be useful to understand the dynamics of spin density in systems with flat bands. 

As is well understood, the Carroll limit implies ultralocality with the collapse of local lightcones, while at the same time its reciprocal nature compared to the Galilean limit imbues it with an ultrarelativistic character. Intuitively, it appears that one is describing the same physics with respect to two different observers. The observer intrinsic to the setup gets restricted to a null hypersurface in the Carroll limit, $c\to 0$, and describes the physics as becoming ultralocal. At the same time, an extrinsic observer, who observes the system as getting boosted closer and closer to the speed of light, $v/c \to 1$, sees it eventually get restricted to evolve along a null hypersurface, and thus becoming Carrollian. It would be worthwhile to put this intuitive picture of Carrollian physics as being both ultralocal and ultrarelativistic, depending upon the choice of observer, on a solid mathematical footing, more generally beyond the observed connections at the level of Carroll physics and ultrarelativistic hydrodynamics.

We leave the above research directions for future explorations.

\acknowledgments
This paper has benefited from discussions with A.~Bagchi, W.~Florkowski, D.~Grumiller, K.~Kolekar, F.~Pe\~{n}a-Benitez, R.~Ryblewski and P.~Sur\'{o}wka. AS also acknowledges helpful discussions with participants of the workshop ``Discussions on Quantum Spacetime'' at the Lodha Mathematical Sciences Institute, Mumbai. AS would like to acknowledge the hospitality of the Institute of Theoretical Physics at the Jagiellonian University, Krak\'{o}w, Poland; the Institute of Theoretical Physics at the Wroc\l{}aw University of Science and Technology, Wroc\l{}aw, Poland; the Department of Physics and Astronomy at the Ghent University, Ghent, Belgium; the Institute for Theoretical Physics at the Technische Universit\"{a}t Wien, Vienna, Austria; and the Tata Institute of Fundamental Research, Mumbai, India, during the course of this work. RS is supported by a postdoctoral fellowship from the West University of Timisoara, Romania. PS is supported by an IIT Kanpur
Institute Assistantship.

\appendix
\section{PUL decomposition of the relativistic hydrodynamic equations}
\label{full_terms}
In this appendix, for the sake of completeness, we present the full PUL decomposition of the relativistic hydrodynamic equations \eqref{hydro_eqs}. Though not utilized in the main text, where we focused only on terms that give non-vanishing contributions in the Carroll limit $c\to 0$, this decomposition is relevant if one needs to compute subleading terms in the hydrodynamic equations in an expansion about $c = 0$, dubbed the ``Carrollian regime'' \cite{deBoer:2023fnj, Kolekar:2024cfg}.  

The PUL decomposition of the four divergence of the energy-momentum tensor is given by,
\begin{equation}
\label{f-PUL-stress-tensor-conservation}
\nabla_\mu T^\mu_{~\,\nu}=\overset{\{0\}}{\nabla_\mu T^\mu_{~\,\nu}}+c^2 \overset{\{2\}}{\nabla_\mu T^\mu_{~\,\nu}}+c^4 \overset{\{4\}}{\nabla_\mu T^\mu_{~\,\nu}},
\end{equation}
where
\begin{subequations}
    \begin{align}
        &\overset{\{0\}}{\nabla_\mu T^\mu_{~\,\nu}}=\widetilde{\N}_\m \overset{(0)}{T}{\!}^\m_{~\n}+\widetilde{\mc K}L_\s \overset{(0)}{T}{\!}^\s_{~\n}-C^\s_{ \m\n}\overset{(0)}{T}{\!}^\m_{~\s}-\overset{(-2)}{\G}{\!\!}^\s_{ \m\n}\overset{(2)}{T}{\!}^\m_{~\s},\\
        &\overset{\{2\}}{\nabla_\mu T^\mu_{~\,\nu}}=\widetilde{\N}_\m \overset{(2)}{T}{\!}^\m_{~\n}+\widetilde{\mc K}L_\s \overset{(2)}{T}{\!}^\s_{~\n}-C^\s_{ \m\n}\overset{(2)}{T}{\!}^\m_{~\s}-\overset{(-2)}{\G}{\!\!}^\s_{ \m\n} \overset{(4)}{T}{\!}^\m_{~\s}-\overset{(2)}{\G}{\!}^\s_{\m\n} \overset{(0)}{T}{\!}^\m_{~\s},\\
        &\overset{\{4\}}{\nabla_\mu T^\mu_{~\,\nu}}=\widetilde{\N}_\m \overset{(4)}{T}{\!}^\m_{~\n}+\widetilde{\mc K}L_\s \overset{(4)}{T}{\!}^\s_{~\n}-C^\s_{ \m\n}\overset{(4)}{T}{\!}^\m_{~\s}-\overset{(2)}{\G}{\!}^\s_{\m\n}\overset{(2)}{T}{\!}^\m_{~\s}.
    \end{align}
\end{subequations}
In writing the above, we have used the PUL decomposition of the energy-momentum tensor, eq.~\eqref{EM_decomp_1}, and the Levi-Civita connection, eq.~\eqref{lcc_decomp}. Now, to obtain the energy and momentum equations, we further project this decomposition along $K^\n$ and $H^{\s\n}$, respectively.

\medskip

\noindent \ding{104} \textit{The energy equation}: We now take the projection of eq.~\eqref{f-PUL-stress-tensor-conservation} along $K^\nu$, which yields the PUL decomposition for the left hand side of the energy equation $K^\n \N_\m T^\m_{~\,\n} = 0$. It reads as

\be
K^\n \N_\m T^\m_{~\,\n}=K^\n \overset{\{0\}}{\nabla_\mu T^\mu_{~\,\nu}}+c^2 K^\n \overset{\{2\}}{\nabla_\mu T^\mu_{~\,\nu}}+c^4 K^\n \overset{\{4\}}{\nabla_\mu T^\mu_{~\,\nu}},
\ee
where the terms at different orders in the PUL decomposition are,
\begin{subequations}
    \begin{align}
        &K^\n \overset{\{0\}}{\nabla_\mu T^\mu_{~\,\nu}}=-K^\m \del_\m \epsilon +\widetilde{\mc{K}} (\epsilon+P),\\
        &K^\n \overset{\{2\}}{\nabla_\mu T^\mu_{~\,\nu}}=-\widetilde{\N}_\m [\mathfrak u^\m (\epsilon+P)]+[L_\n (\pounds_K -2 \widetilde{ \mc K}) +\widetilde{\mc K}_{\m\n}\mathfrak u^\m ](\epsilon+P)\mathfrak u^\n,\\
        &K^\n \overset{\{4\}}{\nabla_\mu T^\mu_{~\,\nu}}=\widetilde{\N}_\m[(\epsilon+P)\mathfrak u^\m \mathfrak u^\s L_\s]+(\epsilon+P)(L_\s \mathfrak u^\s)\big[\widetilde{K}(L_\s \mathfrak u^\s)+\mathfrak u^\lambda \pounds_K L_\lambda\big].
    \end{align}
\end{subequations}

\medskip

\noindent \ding{104} \textit{The momentum equation}: Next, we compute the spatial projection of eq.~\eqref{f-PUL-stress-tensor-conservation} by contracting it with $H^{\s \n}$, which produces the PUL decomposition for the left hand side of the momentum equation $H^{\s\n} \N_\m T^\m_{~\,\n} = 0$. The decomposition reads,
\be
H^{\s\n} \N_\m T^\m_{~\,\n}=H^{\s\n}\overset{\{0\}}{ \N_\m T^\m_{~\,\n}}+c^2 H^{\s\n}\overset{\{2\}}{ \N_\m T^\m_{~\,\n}}+c^4 H^{\s\n} \overset{\{4\}}{\N_\m T^\m_{~\,\n}},
\ee
where
\begin{subequations}
\allowdisplaybreaks
    \begin{align}
        &H^{\s\n}\overset{\{0\}}{ \N_\m T^\m_{~\,\n}}=H^{\s\n}  \big[\del_\n P +(\e+P)  (2K^\rho \del_{[\rho} L_{\n]} - \mf{u}^\rho (\widetilde{\mc{K}} H_{\rho\n} + \widetilde{\mc{K}}_{\rho\n})) \\
        &\qquad\qquad\qquad\qquad+ K^\rho \widetilde{\N}_\rho((\e+P)H_{\n\lambda}\mf{u}^\lambda) \big],\nonumber \\
        &H^{\s\n}\overset{\{2\}}{ \N_\m T^\m_{~\,\n}}=H^{\s\n}\big[\widetilde{\N}_\m ((\epsilon+P)H_{\n\rho}\mathfrak u^\rho \mathfrak u^\m)\\
        &\qquad\qquad\qquad\qquad+(\epsilon+P)\mathfrak u^\alpha\big((\delta^\m_{\ \alpha}-L_\alpha K^\m)(dL)_{\m\n}+\mathfrak{u}^\rho L_\rho(\widetilde{\mc K}H_{\alpha\n}+\widetilde{K}_{\alpha \n})\big)\big],\nonumber\\
        &H^{\s\n}\overset{\{4\}}{ \N_\m T^\m_{~\,\n}}=(\epsilon+P)H^{\s \n}(dL)_{\n\lambda}\mathfrak u^\lambda\mathfrak u^\alpha L_\alpha.
    \end{align}
\end{subequations}
\noindent \ding{104} \textit{The spin current equation}: Finally, we present the PUL decomposition for the left hand side of the spin current equation $\N_\m S^{\m\n\lambda}=0$, given by
\be
\N_\m S^{\m\n\lambda}= \overset{\{0\}}{\N_\m S^{\m\n\lambda}}+c^2  \overset{\{2\}}{\N_\m S^{\m\n\lambda}}+ c^4 \overset{\{4\}}{\N_\m S^{\m\n\lambda}}+c^6 \overset{\{6\}}{\N_\m S^{\m\n\lambda}},
\ee
where
\begin{subequations}
\allowdisplaybreaks
    \begin{align}
        &\overset{\{0\}}{\N_\m S^{\m\n\lambda}}=K^\m \widetilde{\N}_\m {\overset{(0)}{\O}}{\!}^{\n\lambda} - \widetilde{\mc{K}} {\overset{(0)}{\O}}{\!}^{\n\lambda} + 2  \widetilde{\mc{K}}_{\m\s} \mf{u}^\s K^{[\n} {\overset{(0)}{\O}}{\!}^{\lambda]\m},\\
        &\overset{\{2\}}{\N_\m S^{\m\n\lambda}}=K^\m \widetilde{\N}_\m {\overset{(2)}{\O}}{\!}^{\n\lambda} - \widetilde{\mc{K}} {\overset{(2)}{\O}}{\!}^{\n\lambda} + 2  \widetilde{\mc{K}}_{\m\s} \mf{u}^\s K^{[\n} {\overset{(2)}{\O}}{\!}^{\lambda]\m}+\widetilde{\N }_\m[\mathfrak{u}^\m {\overset{(0)}{\O}}{\!}^{\n\lambda}]+\widetilde{\mc K}\mathfrak u^\sigma L_\sigma {\overset{(0)}{\O}}{\!}^{\n\lambda} \\
        &\qquad\qquad~~ +2 H^{\rho [\n}{\overset{(0)}{\O}}{\!}^{\lambda]\s}[ L_{(\m}(dL)_{\s) \rho}K^\m-\widetilde{\mc K}_{\m \rho}\mathfrak{u}^\m L_\s],\nonumber \\
        &\overset{\{4\}}{\N_\m S^{\m\n\lambda}}=\widetilde{\N }_\m[\mathfrak{u}^\m {\overset{(2)}{\O}}{\!}^{\n\lambda}]+\widetilde{\mc K}\mathfrak u^\sigma L_\sigma {\overset{(2)}{\O}}{\!}^{\n\lambda}+2 H^{\rho [\n}{\overset{(2)}{\O}}{\!}^{\lambda]\s}[ L_{(\m}(dL)_{\s) \rho}K^\m-\widetilde{\mc K}_{\m \rho}\mathfrak{u}^\m L_\s]\\
        &\qquad\qquad~~ +2 H^{\rho [\n}{\overset{(0)}{\O}}{\!}^{\lambda]\s} L_{(\m}(dL)_{\s) \rho}\mathfrak{u}^\m,\nonumber\\
        &\overset{\{6\}}{\N_\m S^{\m\n\lambda}}=2 L_{(\m}(dL)_{\s)\rho}\mathfrak u^\m H^{\rho[\n}{\overset{(2)}{\O}}{\!}^{\lambda]\s}.
    \end{align}
\end{subequations}

\section{Local Carroll boost invariance of the hydrodynamic equations}
\label{symmetries}
In this appendix, we prove the local Carroll boost invariance of the Carroll hydrodynamic equations \eqref{carr_hyd_1}, \eqref{carr_hyd_2}, \eqref{carroll_spin_eqn}, and the conformal constraint eq.~\eqref{spin_conf_const}.

\medskip

\noindent \ding{104} \textit{The energy equation}: Consider first the energy equation \eqref{carr_hyd_1}. Let us rewrite it as
\be
k^\m \del_\m \varepsilon - (\varepsilon + p) \widehat{\mc{K}} = 0\,.
\label{CI1}
\ee
Now, following the local Carroll boost transformations in eqs.~\eqref{lcb_trans} and \eqref{lcb_trans_2}, the Carrollian extrinsic curvature $\widehat{\mc{K}}_{\m\n}$ is local Carroll boost invariant. For its trace, one therefore has
\be
\d_\lambda \widehat{\mc{K}} = (\d_\lambda h^{\m\n}) \widehat{\mc{K}}_{\m\n} = (\lambda^\m k^\n + \lambda ^\n k^\m) \widehat{\mc{K}}_{\m\n} = 0\, ,
\ee
as the Carrollian extrinsic curvature is purely spatial in nature, implying $k^\m \widehat{\mc{K}}_{\m\n} = 0$. Thus, the trace of the Carrollian extrinsic curvature is also local Carroll boost invariant. As all other quantities in eq.~\eqref{CI1} are local Carroll boost invariant, this ensures that the energy equation itself remains invariant.

\medskip

\noindent \ding{104} \textit{The momentum equation}: Consider next the momentum eq.~\eqref{carr_hyd_2}, which can be expressed as
\be
h^{\m\n} \del_\n p  + (\varepsilon + p) (\xi^\m - \widehat{\mc{K}} h^{\m\n} \vec{\mf{u}}_\n) + h^{\m\n} k^\s \widehat{\N}_\s ((\varepsilon+p) \vec{\mf{u}}_\n) = 0.
\label{CI2}
\ee
Let us compute the variation of the LHS of eq.~\eqref{CI2} under a local Carroll boost term by term. We have
\be
\d_\lambda\big[h^{\m\n} \del_\n p\big] = (\lambda^\m k^\n + \lambda^\n k^\m) \del_\n p.
\label{var_1}
\ee
One can compute the quantity $\lambda^\n \del_\n p$ by contracting eq.~\eqref{CI2} with $\lambda_\m$. Substituting the resulting expression into eq.~\eqref{var_1} yields
\be
\d_\lambda\big[h^{\m\n} \del_\n p\big] = \lambda^\m k^\n \del_\n p - k^\m\big[(\varepsilon+p)(\lambda_\n \xi^\n - \widehat{\mc{K}} \lambda^\nu \vec{\mf{u}}_\n) + \lambda^\nu k^\s \widehat{\N}_\s((\varepsilon+p)\vec{\mf{u}}_\n)\big].
\label{var_2}
\ee

Next, to compute the variation of the second term on the LHS of eq.~\eqref{CI2} under a local Carroll boost, it is prudent to first compute the variation of $\xi^\m$, which turns out to be
\be
\d_\lambda \xi^\m = k^\m \lambda_\n \xi^\n + h^{\m\n} (\pounds_k\lambda_\n + \lambda^\s \widehat{\mc{K}}_{\s\n}).
\ee
Using this, one gets
\be
\d_\lambda\big[(\varepsilon + p) (\xi^\m - \widehat{\mc{K}} h^{\m\n} \vec{\mf{u}}_\n)\big] = (\varepsilon+p) \big[k^\m \lambda_\n \xi^\n + h^{\m\n} (\pounds_k\lambda_\n + \lambda^\s \widehat{\mc{K}}_{\s\n}) + \widehat{\mc{K}} (\lambda^\m - k^\m \lambda^\n \vec{\mf{u}}_\n)\big] .
\label{var_3}
\ee

Finally, the variation of the last term in eq.~\eqref{CI2} under a local Carroll boost is given by
\be
\begin{split}
\d_\lambda\big[h^{\m\n} k^\s \widehat{\N}_\s ((\varepsilon+p) \vec{\mf{u}}_\n)\big] = &- (\varepsilon+p) \big[h^{\m\n}(\pounds_k \lambda_\n + \lambda^\s \widehat{\mc{K}}_{\s\n}) + \widehat{\mc{K}}\lambda^\m\big] - \lambda^\m k^\n \del_\n p \\
&+ \lambda^\n k^\m k^\s \widehat{\N}_\s((\varepsilon+p) \vec{\mf{u}}_\n) \, ,
\end{split}
\label{var_4}
\ee
where we have made use of the variation of the Carroll compatible connection under a local Carroll boost transformation, eq.~\eqref{conn_var}, as well as the energy equation \eqref{CI1} in simplifying the result. Combining the results obtained in eqs.~\eqref{var_2}, \eqref{var_3} and \eqref{var_4}, it is straightforward to see that the momentum equation \eqref{CI2} is also local Carroll boost invariant.

\medskip

\noindent \ding{104} \textit{The spin current equation}: We now consider the equation for the Carrollian spin current, eq.~\eqref{carroll_spin_eqn}, which can be expressed as 
\be
k^\m \widehat{\N}_\m \mf{s}^{\n\rho} - \widehat{\mc{K}}  \mf{s}^{\n\rho} + 2  \widehat{\mc{K}}_{\m\s} \hat{\mf{u}}^\s k^{[\n} \mf{s}^{\rho]\m} = 0.
\label{CI3}
\ee
The local Carroll boost transformation properties for individual terms in the above equation can straightforwardly be computed to give
\begin{subequations}
\begin{align}
&\d_\lambda\big[k^\m \widehat{\N}_\m \mf{s}^{\n\rho}\big] = 2 \widehat{\mc{K}}_{\m\s} \lambda^\s k^{[\n} \mf{s}^{\rho]\m},\label{var_s_1}\\
&\d_\lambda\big[\widehat{\mc{K}}  \mf{s}^{\n\rho}\big] = 0,\label{var_s_2}\\
&\d_\lambda\big[2  \widehat{\mc{K}}_{\m\s} \hat{\mf{u}}^\s k^{[\n} \mf{s}^{\rho]\m}\big] = - 2 \widehat{\mc{K}}_{\m\s} \lambda^\s k^{[\n} \mf{s}^{\rho]\m},\label{var_s_3}
\end{align}
\end{subequations}
where we have used the fact that the Carrollian spin density $\mf{s}^{\m\n}$ is local Carroll boost invariant, as discussed in the last paragraph of subsection \ref{carroll_spin_hydro}. Combining the results in eqs.~\eqref{var_s_1} - \eqref{var_s_3}, one can conclude that the spin current equation \eqref{CI3} is local Carroll boost invariant.

\medskip

\noindent \ding{104} \textit{The conformal constraint}: Finally, let us consider the conformal constraint eq.~\eqref{spin_conf_const}. The variation of its individual terms under a local Carroll boost transformation is
\begin{subequations}
\begin{align}
&\d_\lambda\big[h^{\m\n}\widehat{\N}_\m \vec{\varphi}_\n\big] = \lambda^\n k^\m \del_\m \vec{\varphi}_\n + \vec{\varphi}_\rho \big[2h^{\rho\nu} \lambda^\m \widehat{\mc{K}}_{\m\n} + \lambda^\m \del_\m k^\rho - \widehat{\mc{K}}\lambda^\rho\big],\label{var_cc_1}\\
&\d_\lambda\big[(k^\m \del_\m-\widehat{\mc{K}}) \hat{\mf{u}}^\nu \vec{\varphi}_\nu\big] = - (k^\m \del_\m-\widehat{\mc{K}}) \lambda^\n \vec{\varphi}_\n\, ,\label{var_cc_2}\\
&\d_\lambda\big[k^\m h^{\n\rho} \vec{\varphi}_\rho (\del_\m \ell_\n - \del_\n \ell_\m)\big] = - \vec{\varphi}_\rho \big[2h^{\rho\nu} \lambda^\m \widehat{\mc{K}}_{\m\n} - \pounds_k\lambda^\rho\big].\label{var_cc_3}
\end{align}
\end{subequations}
It is simple to verify that on combining the results in eqs.~\eqref{var_cc_1} - \eqref{var_cc_3}, the variation of the LHS of eq.~\eqref{spin_conf_const} vanishes, implying that the conformal constraint is also local Carroll boost invariant.

\section{Spin current in the Papapetrou-Randers parametrization}
\label{pr_gauge}
In this appendix, we present the derivation of the equation governing the spin current for an ideal Carroll fluid using the Papapetrou-Randers (PR) parametrization of the background geometry and the fluid degrees of freedom. 

One can express any Lorentzian metric in the PR form as
\be
ds^2 = - c^2 (\O dt - b_i dx^i)^2 + a_{ij} dx^i dx^j,
\label{pr_metric}
\ee
where the PR variables $\O, b_i$ and $a_{ij}$ are functions of the coordinates $(t, x^i)$. 
It is similar in spirit to the celebrated Arnowitt-Deser-Misner (ADM) parametrization \cite {PhysRev.116.1322, Arnowitt:1962hi} for the metric of a Lorentzian spacetime, given by
\be
ds^2 = -c^2 N^2 dt^2 + \g_{ij}\big(dx^i+cN^i dt\big)\big(dx^j+cN^j dt\big),
\label{adm_metric}
\ee
where $N, N^i$ and $\g_{ij}$ are functions of the coordinates $(t,x^i)$. 
Both PR and ADM parametrizations locally decompose the spacetime geometry as $\mathbb{R} \times \Sigma$, with $\mathbb{R}$ denoting the time and $\Sigma$ the spatial directions. 

In the ADM parametrization, spacetime is assumed to be foliated in terms of spacelike hypersurfaces $\Sigma_t$, labeled by the time coordinate $t$, and endowed with the metric $\g_{ij}(t,x^k)$. The lapse and shift, $N$ and $N^i$, respectively, determine how these spacelike hypersurfaces are welded together to produce the full spacetime geometry (left panel of fig.~\ref{fig:adm_pr}). More specifically, the lapse function $N$ determines the proper time $d\tau$ between infinitesimally separated spacelike hypersurfaces $\Sigma_t$ and $\Sigma_{t+dt}$, viz.~$d\tau=Ndt$, for an observer moving orthogonal to the hypersurfaces. This is so because the unit vector normal to the spacelike hypersurface $\Sigma_t$ with the induced metric $\g_{ij}$ is $n^\m =  (1/N, -N^i/N)$. An observer moving along $n^\m$, and thus orthogonal to the hypersurface $\Sigma_t$, satisfies $u^\m \equiv dx^\m/d\tau = c n^\m$, $u^\m u_\m = -c^2$, thereby implying $d\tau = N dt$ (as well as $dx^i/d\tau = - c N^i/N \Rightarrow v^i\equiv dx^i/dt = -cN^i$). On the other hand, the shift $N^i$ determines the relative position of points on the coordinate grids on the said hypersurfaces i.e.~for the observer moving orthogonal to the hypersurfaces, the shift $N^i$ determines what the point labeled $x^i$ on $\Sigma_t$ maps to on $\Sigma_{t+dt}$. Another interpretation of the shift $N^i$ is that it determines the amount by which the coordinate time evolution, generated by $\del_t$, differs from time evolution along the normal, generated by $n^\m \del_\m = N^{-1}(\del_t - N^i \del_i)$. By comparing eqs.~\eqref{pr_metric} and \eqref{adm_metric}, one finds that data for the PR parametrization is related to that of the ADM parametrization via
\be
\O^2 = N^2 - \g_{ij} N^i N^j\, , \quad c\,\O\, b_i = \g_{ij} N^j \, , \quad a_{ij} -c^2 b_i b_j = \g_{ij}\,.
\label{pradmrel1}
\ee

In the PR parametrization eq.~\eqref{pr_metric} (see right panel of fig.~\ref{fig:adm_pr} for a schematic depiction), suppose an observer moves such that $dx^i = 0$. Then one has $d\tau = \Omega dt$, implying that unlike the lapse $N$ in the ADM parametrization which captures the notion of proper time for observers moving orthogonal to the spacelike hypersurfaces, the scale factor $\Omega$ in the PR parametrization determines the proper time for coordinate observers i.e.~observers moving along the integral curves generated by $\del_t$. 
\begin{savenotes}
\begin{figure}[t]
    \centering
    \begin{subfigure}[t]{0.45\textwidth}
        \centering
        \includegraphics[scale=1]{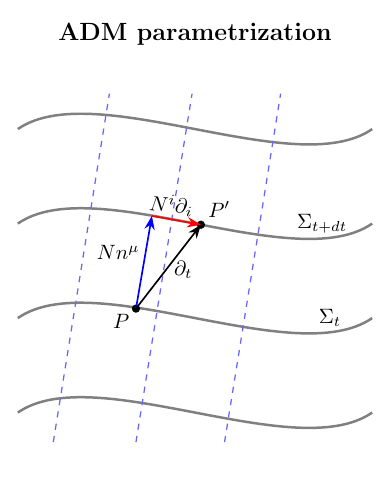}
        \label{fig:pdf1}
    \end{subfigure}
    \hfill
    \begin{subfigure}[t]{0.45\textwidth}
        \centering
        \includegraphics[width=\linewidth]{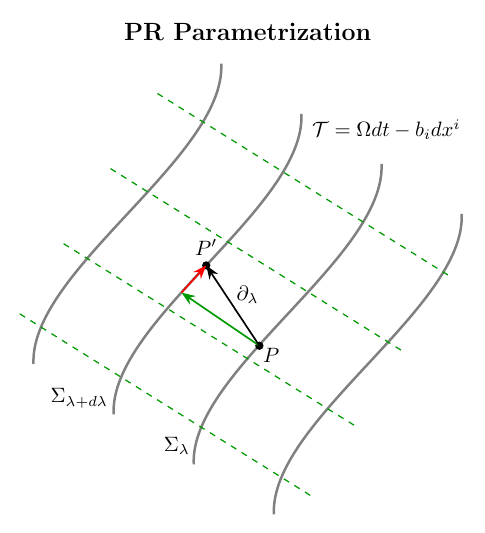}
        \label{fig:pdf2}
    \end{subfigure}
    \captionsetup{justification=RaggedRight,singlelinecheck=false}
    \renewcommand{\footnotelayout}{\justifying}
    \renewcommand{\thempfootnote}{\arabic{mpfootnote}}
    \setcounter{mpfootnote}{\value{footnote}}
    \caption[ADM vs PR]{\small{\underline{Left panel}: A schematic depiction of the ADM parametrization. Spacetime is foliated into spacelike hypersurfaces $\Sigma_t$, labeled by their time coordinate $t$. The blue lines depict a congruence of timelike curves orthogonal to the foliation. Coordinate time evolution $\del_t$ decomposes into a normal and a tangential component, proportional respectively to the lapse $N$ and shift $N^i$.\\ \underline{Right panel}: A schematic depiction of the PR parametrization. The foliation shown assumes that the clock-form $\mathcal{T}$ is integrable i.e.~$\mathcal{T} = f(t,x^i) \, d\lambda$, where $\lambda$ is itself a function of $(t,x^i)$. Equivalently (by the Frobenius theorem), $\mathcal{T}$ is hypersurface orthogonal i.e.~kernel$(\mathcal{T})$\footnote{A one-form $\omega$ acts on vectors $V$ to give real numbers: $\omega(V) \in \mathbb{R}$. The entity kernel($\omega$) is the set of vectors such that $\omega(V) = 0$.} defines the codimension-1 tangent spaces of a foliation by hypersurfaces $\lambda(t,x^i) =$ constant,\footnote{Suppose the clock-form is integrable i.e. $\mathcal{T} = f(t,x^i)\, d\lambda$. For a tangent vector $X$ on the hypersurface $\lambda(t,x^i) =$ constant, one has $d\lambda(X) = 0$. This would mean $\mathcal{T}(X) = 0$ i.e.~every vector in the tangent space belongs to kernel($\mathcal{T}$). Since kernel($\mathcal{T}$) as well as the tangent space are both codimension-1, they coincide, which is the statement of hypersurface orthogonality in our context (as the dual vector $v^\mu \equiv g^{\mu\nu} \mathcal{T}_\nu$ will be orthogonal to constant-$\lambda$ hypersurfaces i.e.~$g(v,X) = 0$).} represented by the gray curves. The green curves depict the integral curves of the timelike vector field dual to $\mathcal{T}$ i.e.~$v^\mu \equiv g^{\m\n} \mathcal{T}_\n$, which is orthogonal to the constant-$\lambda$ hypersurfaces and defines a congruence of observers. The tangential shift (red arrow) measures the displacement on a constant-$\lambda$ hypersurface between the point reached by evolving along the observer congruence and the point reached by the chosen coordinate evolution $\del_\lambda$. Note that although PR parametrization is fully general, the assumption of integrability/hypersurface orthogonality of the clock-form is needed to arrive at the foliation above. Without this, while the congruence of timelike observers $v^\mu$ will remain well defined, the codimension-1 hyperplanes defined by kernel($\mathcal{T})$ at different points need not fit together to form a smooth foliation by equal-time hypersurfaces.}}
    \label{fig:adm_pr}
\end{figure}
\end{savenotes}
\addtocounter{footnote}{2}
\hspace{-4mm} Next, consider another observer moving with the coordinate velocity $v^i \equiv dx^i/dt$. For such an observer one gets,
\be
d\tau = dt \sqrt{\big(\Omega - b_i v^i\big)^2 - \f{1}{c^2} a_{ij}v^i v^j}.
\ee
While the term $a_{ij}v^i v^j/c^2$ captures the purely kinematic time dilation effect associated with spatial motion,\footnote{This is so because even in flat Minkowski spacetime, with $ds^2 = - c^2 dt^2 + \d_{ij} dx^i dx^j$, which consequently has $\Omega = 1, b_i = 0, a_{ij} = \d_{ij}$, the proper time contains the kinematic time dilation effect due to $a_{ij}$, viz.~$d\tau = dt \sqrt{1- \f{1}{c^2}\d_{ij} v^i v^j}$, for observers moving with the coordinate velocity $v^i \equiv dx^i/dt$.} the object $b_i$ encapsulates a genuine geometric feature of the PR parametrization. It encodes the local intertwining of temporal and spatial directions that gives an additional velocity-dependent contribution to the observer's proper time. Interestingly, the term linear in $b_i$ appearing in $d\tau^2$ is an odd function of the observer's velocity $v^i$, unlike the purely kinematic contribution, which is an even function of $v^i$. Thus, for non-zero $b_i$, the proper time for a moving observer acquires a directional dependence in the PR parametrization. Further, the induced metric on a constant-$t$ spacelike hypersurface in the PR parametrization is $a_{ij} - c^2 b_i b_j$. Thus, in contrast to the ADM parametrization, where the fundamental spatial tensor $\g_{ij}$ is the induced metric on spacelike hypersurfaces, in the PR parametrization $a_{ij}$ is not the induced metric on a constant-$t$ hypersurface. Instead, the induced metric on a constant-$t$ spacelike hypersurface is determined jointly by $a_{ij}$ and $b_i$.  

From a geometric perspective, in the ADM parametrization, the starting point is the choice of a foliation of spacetime in terms of spacelike hypersurfaces, with the spatial metric on these hypersurfaces playing the fundamental role. In appropriately chosen coordinates, the spatial metric is $\g_{ij}(t,x^k)$, with $t$ labeling the hypersurfaces, while the lapse and shift fields $(N, N^i)$ are then introduced to construct the full spacetime metric, eq.~\eqref{adm_metric}. Contrary to this, in the PR parametrization, the starting point is the choice of a nowhere-vanishing timelike one-form, the clock-form $\mathcal{T}$, which plays the fundamental role by defining a local notion of time. In appropriately chosen coordinates the clock-form can be written as $\mathcal{T} = \O dt - b_i dx^i$, eq.~\eqref{pr_metric}. On top of this, a positive-definite symmetric tensor $a_{ij}$ is additionally introduced to complete the parametrization of the full spacetime metric. Thus, in PR parametrization the metric takes the form $ds^2 = -c^2 \mathcal{T}^2 + a_{ij} dx^i dx^j$. This gives a nice geometric interpretation to $b_i$ and $a_{ij}$: the positive-definite symmetric tensor $a_{ij}$ is the metric that measures spatial distances with respect to the observer adapted to the clock-form $\mathcal{T}$, while $b_i$ measures how spatial displacements affect the local notion of time for such an observer.\footnote{Said differently, $b_i$ measures the failure of the clock-form to coincide with the coordinate time one-form $dt$. Analogously, in the ADM parametrization, the shift $N^i$ measures the failure of the coordinate time evolution vector $\del_t$ to be orthogonal to the spacelike foliation.} Both ADM and PR parametrizations are general enough and therefore the metric of any Lorentzian spacetime can locally be written in either form. What they differ in is their complementary geometric viewpoints: while the ADM parametrization views spacetime as first foliated in terms of spacelike hypersurfaces carrying a spatial metric and introduces additional structure in the form of lapse and shift fields to reconstruct the full geometry, in the PR parametrization one first chooses a local notion of time carried by the clock-form and then introduces additional structure in the form of a positive-definite tensor to reconstruct the full spacetime metric. In other words, the primary geometric object in the ADM parametrization is the spatial metric, while in PR parametrization it is the clock-form. The ADM parametrization is well suited for performing a Hamiltonian analysis, while the PR parametrization is naturally geared towards taking the Carroll limit, which will be evident from the discussion below.

From the PUL perspective, by comparing with eq.~\eqref{metric_pul}, the choice of PR parametrization eq.~\eqref{pr_metric} for the background geometry corresponds to 
\be
\begin{split}
&K^\m \del_\m = \f{1}{\O} \del_t, \qquad\qquad\quad~ H_{\m\n} dx^\m dx^\n =  a_{ij} dx^i dx^j, \\ &L_\m dx^\m = - \O dt + b_i dx^i, \quad H^{\m\n} \del_\m \del_\n = \f{b^2}{\O^2} \del_t^2 + \f{2b^i}{\O} \del_t \del_i + a^{ij} \del_i \del_j,
\end{split}
\ee
where $b^2 \equiv b^i b_i, b^i \equiv a^{ij} b_j$, with $a^{ij}$ being the inverse of $a_{ij}$ i.e.~$a^{ij} a_{jk} = \d^i_k$. Then, on taking the $c\to 0$ limit, one lands on a Carrollian structure with the degenerate spatial metric $h_{\m\n}$, its kernel $k^\m$, the clock-form $\ell_\m$ and the co-metric $h^{\m\n}$ given by
\be
\begin{split}
&k^\m \del_\m = \f{1}{\O} \del_t, \qquad\qquad\quad~ h_{\m\n} dx^\m dx^\n =  a_{ij} dx^i dx^j, \\ 
&\ell_\m dx^\m = - \O dt + b_i dx^i, \quad h^{\m\n} \del_\m \del_\n = \f{b^2}{\O^2} \del_t^2 + \f{2b^i}{\O} \del_t \del_i + a^{ij} \del_i \del_j.
\end{split}
\label{map_pul_pr}
\ee
From the above, it is clear that PR parametrization performs a $3+1$ split of the background Lorentzian geometry in such a way that arriving at a Carroll structure in the $c\to 0$ limit becomes straightforward. Note that while the general PUL decomposition eq.~\eqref{metric_pul} allows one to include subleading terms in a $c\to 0$ expansion of the geometry, eq.~\eqref{PUL_Carr_Exp}, the PR parametrization captures only the leading Carroll structure that emerges in this limit. This is so because the geometric objects in PR parametrization viz.~$\O, b_i, a_{ij}$ are assumed to be rigid and do not admit a $c\to 0$ expansion, as opposed to the PUL variables appearing in eq.~\eqref{PUL_Carr_Exp}.

The fluid four-velocity can be parametrized in the PR form as $u^\m \del_\m = \g \del_t + \g v^i \del_i$, with
\be
\g = \f{1+c^2 \vec{\b}\cdot\vec{b}}{\O\sqrt{1-c^2 \b^2}}\, , \quad v^i = \f{c^2 \O \b^i}{1+c^2 \vec{\b}\cdot\vec{b}}\,,
\label{PR_velo_decomp}
\ee
where $\b_i \equiv a_{ij} \b^j$, $\b^2 \equiv \b^i \b_i$, and $\vec{\b}\cdot\vec{b} \equiv \b^i b_i = \b_i b^i$. The field $\b^i$ captures the three degrees of freedom in $u^\m$.\footnote{The field $\b^i$ is the Goldstone boson associated with the spontaneous breaking of local Carroll boost invariance in the thermal state \cite{Armas:2023dcz}. This is an important aspect on which Carroll hydrodynamics differs from relativistic hydrodynamics. In the thermal state, local Lorentz boost invariance is also broken spontaneously in relativistic hydrodynamics. However, the associated Goldstone boson turns out to be proportional to the fluid four-velocity \cite{Nicolis:2015sra, Bagchi:2025vri}, and therefore does not appear independently in the hydrodynamic equations, unlike for Carroll hydrodynamics.} Once again, in terms of the PUL decomposition eq.~\eqref{u_pul}, the PR parametrization of the fluid four-velocity corresponds to the choice
\be
\mf{u}^{\m} \del_\m = \f{1}{c^2\O}\left(\f{1+c^2 \vec{\b}\cdot\vec{b}}{\sqrt{1-c^2 \b^2}}-1\right)\del_t + \f{\b^i}{\sqrt{1-c^2 \b^2}}\del_i\, .
\ee
Unlike the background geometry, the PR parametrized fluid four-velocity admits a $c\to 0$ expansion, including subleading terms \cite{Kolekar:2024cfg}. For our purpose here, the leading terms in this expansion suffice, which are
\be
u^\m\del_\m = \f{1}{\O} \del_t + \mc{O}(c^2).
\ee
For eq.~\eqref{PUL_carr_velo}, this implies that
\be
\hat{\mf{u}}^\m \del_\m = \frac{1}{\Omega}\!\left(\frac{\beta^2}{2}+\vec \beta\cdot \vec b \right)\!\del_t + \b^i \del_i\, , \quad \vec{u}_\m dx^\m = \b_i dx^i.
\label{map_pul_pr_velo}
\ee

The PR parametrization of the background geometry and the fluid four-velocity can now be used to compute the Carroll limit $c\to 0$ of the relativistic hydrodynamic equations, $u^\n \N_\m T^{\m}_{~\,\n} = 0$ and $\Delta^{\s\n} \N_\m T^{\m}_{~\,\n} = 0$, to get
\begin{subequations}
\begin{align}
&\hat{\del}_t \varepsilon = - \mathscr{U} (\varepsilon + p),\label{pr_carr_hyd_1}\\
&\hat{\del}_i p = -\mathscr{V}_i (\varepsilon+p) - (\hat{\del}_t + \mathscr{U})((\varepsilon+p)\b_i).\label{pr_carr_hyd_2}
\end{align}
\label{pr_carr_hyd}
\end{subequations}
Here we have made use of the notation
\be
\mathscr{U} \equiv \f{1}{\O} \del_t \log\sqrt{a}\, , \quad \mathscr{V}_i \equiv \f{1}{\O} (\del_i\O+\del_t b_i),
\ee
where $a\equiv {\rm det}(a_{ij})$. The objects $\mathscr{U}, \mathscr{V}_i$ are respectively known as ``Carrollian expansion'' and ``Carrollian acceleration.'' Further, we have defined
\be
\hat{\del}_t \equiv \f{1}{\O} \del_t \, , \quad \hat{\del}_i \equiv \del_ i + \f{b_i}{\O} \del_t\, .
\label{carr_cov_defs}
\ee
The advantage of working with $\hat{\del}_t$, $\hat{\del}_i$ is that they transform covariantly under the following reduced set of diffeomorphisms,
\be
t \rightarrow t'(t, x^i), \quad x^i \rightarrow x'{}^i(x).
\label{carr_diffeos}
\ee
This subset of the most general possible diffeomorphisms are sometimes referred to as the ``Carroll diffeomorphisms,'' and have the interesting feature of preserving the PR form of the line element eq.~\eqref{pr_metric}. In particular, under eq.~\eqref{carr_diffeos}, one has
\be
\O\rightarrow \O' = \f{\O}{J}\, , \quad a_{ij} \rightarrow a'_{ij} = (J^{-1})_i^{~k} (J^{-1})_j^{~l} a_{kl}\, , \quad b_i \rightarrow b'_{i} = \left(b_k+\f{\O}{J} J_k\right)  (J^{-1})_i^{~k},
\ee
where the Jacobian factors are
\be
J = \f{\del t'}{\del t}\, , \quad J_i = \f{\del t'}{\del x^i}\, , \quad J_i^{~k} = \f{\del x'{}^{k}}{\del x^i}.
\ee
Thus, $\O$ transforms like a scalar density, $a_{ij}$ transforms like a rank-two covariant (spatial) tensor, and $b_i$ transforms like a connection\footnote{After taking the Carroll limit in PR parametrization, $b_i$ is referred to as the ``Ehresmann connection'' on the emergent Carrollian structure, and appears inside the clock-form $\ell_\m dx^\mu$, eq.~\eqref{map_pul_pr}.} under eq.~\eqref{carr_diffeos}. Further, the derivatives in eq.~\eqref{carr_cov_defs} transform via $\hat{\del}'_{t'} = \hat{\del}_t$ and $\hat{\del}'_{i'} = (J^{-1})_i^{~k} \hat{\del}_k$ i.e.~like a scalar and a (spatial) one-form, respectively. Eqs.~\eqref{pr_carr_hyd_1} and \eqref{pr_carr_hyd_2} are the equations of Carroll hydrodynamics for an ideal Carroll fluid written using the PR parametrization, and have been the subject of discussion in \cite{Ciambelli:2018xat, Petkou:2022bmz, Bagchi:2023ysc, Bagchi:2023rwd, Kolekar:2024cfg, Bagchi:2025vri}. With the advent of the PUL parametrization discussed in subsection \ref{pul_setup}, and the identification between the PUL variables and those of the PR parametrization, eqs.~\eqref{map_pul_pr} and \eqref{map_pul_pr_velo}, eqs.~\eqref{pr_carr_hyd_1} and \eqref{pr_carr_hyd_2} can directly be obtained from eqs.~\eqref{carr_hyd_1} and \eqref{carr_hyd_2}, respectively.

Next, let us consider the spin degrees of freedom $\P_\m, \Pi_\m$, which  satisfy the orthogonality conditions $u^\mu \P_\m = u^\m \Pi_\m = 0$. Using the PR parametrization of the fluid four-velocity, eq.~\eqref{PR_velo_decomp}, one finds that these orthogonality conditions imply $\P_t = - v^i \Phi_i$ and $\Pi_t = - v^i \Pi_i$. Thus, in the Carroll limit $c\to 0$, one gets $\Phi_t = \mc{O}(c^2), \,\Pi_t = \mc{O}(c^2)$. The orthogonality conditions can also be written as $u_\m \P^\m = u_\m \Pi^\m = 0$,\footnote{We have {$u_\m dx^\mu =\f{c^2}{\sqrt{1-c^2\beta^2}}(-\O dt + (b_i+\b_i) dx^i)$.}} and can then be solved to compute $\Phi^t, \Pi^t$ in terms of $\Phi^i, \Pi^i$ to get
\be
\Phi^t = \f{1}{\O} (b_i+\b_i) \Phi^i, \quad \Pi^t = \f{1}{\O} (b_i+\b_i) \Pi^i.
\label{ppup_1}
\ee
Now, we have the relations 
\be
\begin{split}
&\Phi_i = a_{ij}\Phi^j + c^2 b_i\b_j \Phi^j\, , \qquad\! \Pi_i = a_{ij}\Pi^j + c^2 b_i\b_j \Pi^j,\\
&\Phi^i = a^{ij}\Phi_j - \f{c^2b^i\b^j\Phi_j}{1+c^2 \vec{\b}\cdot\vec{b}}\, , \quad \Pi^i = a^{ij}\Pi_j - \f{c^2b^i\b^j\Pi_j}{1+c^2 \vec{\b}\cdot\vec{b}}\,.
\end{split}
\ee
In the Carroll limit $c\to 0$, one thus gets $\Phi_i = a_{ij} \Phi^j$, $\Phi^i = a^{ij} \Phi_j$, $\Pi_i = a_{ij} \Pi^j$, and $\Pi^i = a^{ij} \Pi_j$, up to $\mc{O}(c^2)$ corrections. Eq.~\eqref{ppup_1} can thus be written in the Carroll limit as
\be
\Phi^t = \f{1}{\O} (b^i+\b^i) \Phi_i + \mc{O}(c^2), \quad \Pi^t = \f{1}{\O} (b^i+\b^i) \Pi_i + \mc{O}(c^2).
\label{ppup_2}
\ee
Thus, in the Carroll limit, the spin degrees of freedom are encoded in $\Phi_i, \Pi_i$, which can be identified with $\vec{\varphi}_i, \vec{\pi}_i$ that arise from the Carroll limit in the PUL approach, eq.~\eqref{PUL_spin_data}, with $\vec{\varphi}_t = \vec{\pi}_t = 0$ by virtue of eqs.~\eqref{ortho_conds} and \eqref{map_pul_pr}. In the Carroll limit, the non-vanishing components of the spin density $\O^{\m\n}$, eq.~\eqref{omega_decomp}, can then be written in the PR parametrization as
\be
\O^{ti} = - \O^{it} = \f{1}{\O}\left(a^{ij} \vec{\varphi}_j + \f{\varepsilon^{tijk}}{\sqrt{a}}(b_j+\b_j)\vec{\pi}_k\right)+\mc{O}(c^2), \quad \O^{ij} = - \f{\varepsilon^{tijk}}{\sqrt{a}}\vec{\pi}_k + \mc{O}(c^2),
\ee
where we have used $\sqrt{-g} = c\O\sqrt{a}$. Using the above, one can compute the components of the spin current $S^{\m\n\lambda} = u^\m \O^{\n\lambda}$ in the PR parametrization in the $c\to 0$ limit to be
\be
\begin{split}
S^{tti} &= \f{1}{\O^2}\left(a^{ij} \vec{\varphi}_j + \f{\varepsilon^{tijk}}{\sqrt{a}}(b_j+\b_j)\vec{\pi}_k\right)+\mc{O}(c^2),\\
S^{tij} &= - \f{\varepsilon^{tijk}}{\O\sqrt{a}}\vec{\pi}_k + \mc{O}(c^2),\\
S^{ijt} &= \mc{O}(c^2),\\
S^{ijk} &= - c^2 \f{\b^i  \varepsilon^{tjkl}}{\sqrt{a}}\vec{\pi}_l + \mc{O}(c^4).
\end{split}
\ee
Finally, the spin current conservation equation $\N_\m S^{\m\n\lambda} = 0$ in the limit $c\to 0$ in the PR parametrization becomes\footnote{See for instance eq.~(16) of \cite{Kolekar:2024cfg} for the components of the Levi-Civita connection expressed in the PR parametrization, which have been utilized to obtain eq. \eqref{spin_eqs_PR}.}
\begin{subequations}
\begin{align}
\lim_{c\to 0} \N_\m S^{\m ti} &= 0 \Rightarrow (\hat{\del}_t + \mathscr{U}) \mc{X}^i + \hat{\g}^i_{~j} \mc{X}^j + \big(\del_j\O + \O\hat{\g}_{jk}(b^k+\b^k)\big)\mc{Y}^{ij}= 0,\\
\lim_{c\to 0} \N_\m S^{\m ij} &= 0 \Rightarrow (\hat{\del}_t + \mathscr{U})\O\mathcal{Y}^{ij} + 2\O \hat{\g}^{[i}_{~k} \mathcal{Y}^{kj]} = 0. 
\end{align}
\label{spin_eqs_PR}
\end{subequations}
Here we have used the notation
\be
\hat{\g}_{ij} \equiv \f{1}{2} \hat{\del}_t a_{ij} \, , \quad \mathcal{X}^i \equiv a^{ij} \vec{\varphi}_j + \f{\varepsilon^{tijk}}{\sqrt{a}}(b_j+\b_j)\vec{\pi}_k\, ,\quad
\mathcal{Y}^{ij} \equiv  \f{\varepsilon^{tijk}}{\O\sqrt{a}}\vec{\pi}_k\, .
\ee
Note that $\hat{\g}_{ij}$ transforms like a tensor under Carroll diffeomorphisms, eq.~\eqref{carr_diffeos}, and thus its indices can be raised by using the inverse spatial metric i.e.~$\hat{\g}^i_{~j} = a^{ik} \hat{\g}_{kj}$. Further, if the fluid is conformal, one has $\N_\m S_\n^{~\,\n\m} = 0$, or equivalently $\N_\m \Phi^\m = 0$. This condition in the PR parametrization when $c\to 0$ becomes
\be
\hat \partial_i(a^{ij}\vec\varphi_j)+(\hat\partial_t+\mathscr{U})\beta^i\vec{\varphi}_i + \big(\mathscr{V}_i+\hat\gamma^k_{ki}\big) a^{ij}\vec{\varphi}_j = 0,
\label{conf_cond_pr}
\ee
where $\hat{\g}^i_{jk} \equiv \f{1}{2} {a}^{il}(\hat{\del}_j a_{kl} + \hat{\del}_k a_{jl} - \hat{\del}_l a_{jk})$ is referred to as the spatial Levi-Civita-Carroll connection, which actually transforms like a tensor under Carroll diffeomorphisms eq.~\eqref{carr_diffeos}. Eqs.~\eqref{spin_eqs_PR} and \eqref{conf_cond_pr} for the spin current of an ideal Carroll fluid expressed in PR parametrization are some of the key results of the present work. An independent way to arrive at these equations is to start with eqs.~\eqref{carroll_spin_eqn_2} and \eqref{spin_conf_const_2} obtained using PUL parametrization, and specialize to the choice eqs.~\eqref{map_pul_pr} and \eqref{map_pul_pr_velo}.

\subsection{Maps to Bjorken and Gubser flow}
\label{the_maps_pr}
The geometric choices that map the equations of Carroll hydrodynamics in PR parametrization to Bjorken and Gubser flow were worked out in \cite{Bagchi:2023ysc, Bagchi:2023rwd}. To map the Carroll hydrodynamic equations \eqref{pr_carr_hyd} to Bjorken flow eq.~\eqref{Bjorken_eq_1}, along with the associated phenomenological assumptions, one has to choose the data \cite{Bagchi:2023ysc}
\be
\O = 1, \quad b_i +\beta_i = 0, \quad a_{ij}dx^i dx^j = \t^2 d\rho^2 + dx^2 + dy^2,
\label{bj_pr_data}
\ee
while the mapping to Gubser flow on the global dS$_3\times\mathbb{R}$ background eq.~\eqref{Gubser_eq_1}, with the associated phenomenological assumptions, works out for
\be
\O = 1, \quad b_i + \beta_i = 0, \quad a_{ij} dx^i dx^j = \cosh^2\varsigma \, (d\psi^2 + \sin^2\psi\,d\phi^2) + d\rho^2.
\label{gb_pr_data}
\ee
The choices in eq.~\eqref{bj_pr_data} and \eqref{gb_pr_data} can also be seen to follow from eqs.~\eqref{bj_data} and \eqref{gb_data} with the identifications in eqs.~\eqref{map_pul_pr} and \eqref{map_pul_pr_velo}. Further, to arrive at the equations for the spin degrees of freedom in Bjorken and Gubser flow, eqs.~\eqref{bj_spin_eqs_1} and \eqref{gubser_spin_dofs}, respectively, one can has to make the following choices for the PR data in eqs.~\eqref{spin_eqs_PR},
\begin{subequations}
\begin{align}
&{\rm Bjorken}: \vec{\varphi}_i dx^i = \t \hat{f}(\t) d\rho\, , \quad~ \vec{\pi}_i dx^i = \t \hat{g}(\t) d\rho.\\
&{\rm Gubser}: ~\vec{\varphi}_i dx^i = \hat{v}(\varsigma) d\rho \, , \qquad \vec{\pi}_i dx^i = \,\hat{w}(\varsigma) d\rho .
\end{align}
\label{pr_to_spin}
\end{subequations}
These choices also follow from the corresponding ones in the PUL approach, eqs.~\eqref{pul_bj_spin_data} and \eqref{gb_data}, by substituting $\vec{\varphi}_t = \vec{\pi}_t = 0$ valid for the PR parametrization. Note that for the case of Gubser flow, the conformal constraint eq.~\eqref{conf_cond_pr} is trivially satisfied without imposing any additional conditions on $(\hat{v}, \hat{w})$. To summarize, the equations of Carroll hydrodynamics with a spin current in the PR parametrization map to the equations and phenomenological assumptions for Bjorken and Gubser flow, with spin, by enlarging the maps obtained in \cite{Bagchi:2023ysc, Bagchi:2023rwd}, eqs.~\eqref{bj_pr_data} and \eqref{gb_pr_data}, with eq.~\eqref{pr_to_spin}. 

\section{Pseudogauge freedom in Carroll spin hydrodynamics}
\label{carr_pseudo}
In this appendix, we comment upon the pseudogauge freedom in Carroll spin hydrodynamics. In relativistic spin hydrodynamics, as alluded to in section \ref{rel_hydro}, pseudogauge freedom \cite{Hehl:1976vr,Becattini:2011ev,Speranza:2020ilk,Singh:2024qvg,armas2026thermodynamicsidealspinfluids} corresponds to the ambiguity in decomposing the total angular momentum current into an orbital and a spin part. Working in flat Minkowski spacetime, one can shift the energy-momentum tensor and the spin current following eq.~\eqref{pseudogauge}, which leaves the form of the equations of motion, eqs.~\eqref{cons_eq_1} and \eqref{spin_non_cons}, as well as the total angular momentum current eq.~\eqref{total_ang_mom} unchanged (up to total derivative terms). It turns out that analogous to the relativistic case, for a Carroll fluid endowed with a spin current and put on a flat Carroll background, the energy-momentum and spin currents can be redefined leaving the form of their conservation equations as well as the Carrollian angular momentum current unaffected (up to total derivative terms), thus manifesting pseudogauge freedom for Carroll spin hydrodynamics. To wit, we first specialize the background geometry satisfying eq.~\eqref{completeness_2} to the flat Carroll background, which can be obtained by setting\footnote{Note that the flat Carroll background considered here corresponds to the \textit{weak} notion of flatness (see \cite{Duval:2014lpa}). Consequently, due to supertranslations, its isometry algebra is infinite-dimensional, with the finite-dimensional Carroll algebra appearing as a subalgebra. Further, we keep $b_i$ non-zero, as any conclusion drawn about the pseudogauge ambiguity with $b_i$ arbitrary will be invariant under a local Carroll boost. 
}
\begin{equation}\label{flat-data}
    \begin{split}
          &h_{\m\n}dx^\m dx^\n=\delta_{ij}dx^i dx^j,\quad k^\m\del_\mu=\del_t\, ,\\
          &\ell_\m dx^\m=-dt+b_idx^i,\quad h^{\m\n}\del_\m \del_\n=b^2\del_t^2+2b^i\del_i \del_t+\delta^{ij}\del_i\del_j\,.
    \end{split}
\end{equation}
The equations of Carroll spin hydrodynamics, eqs.~\eqref{carr_hyd_pul} and \eqref{carroll_spin_eqn}, on the flat Carroll background take the form
\begin{equation}\label{flat-carroll-fluid-all}
    \del_t \varepsilon=0,\quad \del_ip=-\del_t[(\varepsilon+p)(b_i+\beta_i)],~ \text{and}~~ \del_t \mathfrak{s}^{\m\n}=0,
\end{equation}
where we have used the notation $\vec{\mathfrak{u}}_\m dx^\m\equiv\beta_i dx^i$. Also, the $c\to0$ limit of the relativistic angular momentum current eq.~\eqref{total_ang_mom} can be used to compute the corresponding Carrollian current, denoted by $\widehat J^\m_{\ \n\lambda}$, and given by
\begin{equation}\label{Carroll-tot-ang}
     \widehat{ J}^\m_{\ \alpha\beta} =h_{\alpha \sigma}x^{\sigma}\left[(\varepsilon+p)k^\mu (\ell_\beta+\vec{\mathfrak{u}}_\beta )+p\delta^\mu_{\ \beta}\right]-h_{\beta\sigma}x^{\sigma}\left[(\varepsilon+p)k^\mu (\ell_\alpha+\vec{\mathfrak{u}}_\alpha )+p\delta^\mu_{\ \alpha}\right]-k^\mu \mathfrak{s}^{\sigma \rho}h_{\sigma \alpha}h_{\rho\beta}.
\end{equation}
Substituting eq.~\eqref{flat-data} into eq.~\eqref{Carroll-tot-ang}, one gets the following components for the Carrollian angular momentum current on the flat Carroll background,
\begin{equation}
    \begin{split}
        &\widehat{ J}^t_{\ ti}=x_i\varepsilon \equiv -x_i \mathcal{T}^t_{~\,t},\quad  \widehat{ J}^k_{\ ij}=(x_i \delta^k_{\ j}-x_j \delta^k_{\ i})p\equiv x_i\mathcal{T}^k_{~\,j}-x_j\mathcal{T}^k_{~\,i}\,,\\
        &\widehat{ J}^k_{\ ti}=0,\quad \widehat{ J}^t_{\ ij}=(x_i \delta^k_{\ j}-x_j \delta^k_{\ i})(b_k+\beta_k)(\varepsilon+p)-\mathfrak{s}_{ij}\equiv (x_i \mathcal{T}^t_{~\,j}-x_j  \mathcal{T}^t_{~\,i})-\mathfrak{s}_{ij}\,,
    \end{split}
\end{equation}
where $\mathfrak{s}_{ij}=\delta_{ik}\delta_{jl}\mathfrak{s}^{kl}$, while $\mathcal{T}^\m_{~\,\n}$ denote the components of the Carrollian energy-momentum tensor. It is then straightforward to see that under the following shift 
\be
\mathcal{T}^t_{~\,i}\to \mathcal{T}'^t_{~~\,i}=\mathcal{T}^t_{~\,i}+\frac{1}{2}\del_k Z^k_{~\,i}, \quad \mathfrak{s}_{ij}\to \mathfrak{s}'_{ij}=\mathfrak{s}_{ij}-Z_{ij},
\label{carr_pg}
\ee
where $Z_{ij} \equiv Z_{ij}(x^k)$ is arbitrary, eqs.~\eqref{flat-carroll-fluid-all} remain invariant, while the Carrollian angular momentum current shifts by a total derivative term, 
\begin{equation}
    \widehat{J}^t_{\ ij}\to\widehat{J}'^t_{\ ij}=\widehat{J}^t_{\ ij}+\del_k \left(x_{[i}Z^k_{\ j]} \right).
\end{equation}
This can be immediately understood as the manifestation of the pseudogauge freedom for Carroll spin hydrodynamics. Note that the components $\mathcal{T}^t_{~\,t}, \mathcal{T}^i_{~\,j}$ are not affected by the pseudogauge freedom eq.~\eqref{carr_pg}, while the $\mathfrak{s}^{ti}$ components of Carrollian spin density do not affect the angular momentum current at all, and can thus be shifted around arbitrarily via 
\begin{equation}
\mathfrak{s}^{ti}(x^i)\to\mathfrak{s}'^{ti}(x^i)=\mathfrak{s}^{ti}-Z^{ti}(x^i).
\end{equation}
One can thus choose a distinguished pseudogauge analogous to the Belinfante-Rosenfeld pseudogauge for the relativistic case, given by the choice $Z^{ti}=\mathfrak{s}^{ti}, Z^{ij}=\mathfrak{s}^{ij}$, wherein the Carrollian spin current vanishes identically.

\bibliography{refs}

\end{document}